\documentclass[twocolumn,showpacs,aps,epsfig]{revtex4}

\usepackage{graphicx}
\usepackage{epstopdf}
\usepackage{latexsym}

\usepackage[center]{subfigure}

\begin{document}

 \newcommand{\bq}{\begin{equation}}
 \newcommand{\eq}{\end{equation}}
 \newcommand{\bqn}{\begin{eqnarray}}
 \newcommand{\eqn}{\end{eqnarray}}
 \newcommand{\bqs}{\begin{equation}\begin{split}}
 \newcommand{\eqs}{\end{split}\end{equation}}
 \newcommand{\nb}{\nonumber}
 \newcommand{\lb}{\label}
\newcommand{\PRL}{Phys. Rev. Lett.}
\newcommand{\PL}{Phys. Lett.}
\newcommand{\PR}{Phys. Rev.}
\newcommand{\CQG}{Class. Quantum Grav.}

\title{Branes  in  the $M_{D} \times M_{d^{+}} \times M_{d^{-}}$ Compactification of type II string  on $S^{1}/Z_{2}$ 
and their cosmological applications}
\author{Michael Devin ${}^{1}$}   
\email{Michael_Devin@baylor.edu} 
\author{Tibra Ali ${}^{1}$}   
\email{Tibra_Ali@baylor.edu} 
\author{Gerald Cleaver ${}^{1}$}   
\email{Gerald_Cleaver@baylor.edu}
\author{Anzhong Wang ${}^{1}$}
\email{Anzhong_Wang@baylor.edu}  

\author{Qiang Wu ${}^{1, 2}$}
\email{Qiang_Wu@baylor.edu} 
 
\affiliation{ ${}^{1}$ CASPER, Department of Physics, Baylor University,
Waco, Texas 76798-7316\\
${}^{2}$   Department of Physics, Zhejiang University of
Technology,
Hangzhou 310032,  China }
 
\date{\today}

\begin{abstract}

 In this paper, we study the implementation of brane worlds in type II string 
 theory. Starting with the NS/NS sector of type II string, 
 we first compactify the $\left(D+d_{+} + d_{-}\right)$-dimensional spacetime,
 and reduce the corresponding action to a D-dimensional effective action, where
 the topologies of $M_{d_{+}}$ and $M_{d_{-}}$ are arbitrary. We further
 compactify one of the $(D-1)$ spatial dimensions on an $S^{1}/Z_{2}$ orbifold,
 and derive the gravitational and matter field equations both in the bulk and
 on the branes. Then, we investigate two key issues in such a setup: (i) the
 radion stability and radion mass; and (ii) the localization of gravity, and the
 corresponding Kaluza-Klein (KK) modes. We show explicitly that the radion is stable
 and its mass can be in the order of $GeV$. In addition, the gravity is localized
 on the visible brane, and its spectrum of the gravitational KK towers is discrete 
 and can have a mass gap of $TeV$, too.  The high order Yukawa corrections to the
 4-dimensional Newtonian potential is exponentially suppressed, and can be 
 negligible. Applying such a setup to cosmology, we obtain explicitly the field 
 equations in the bulk and the generalized Friedmann equations on the branes.

\end{abstract}
\preprint{arXiv: xxxxxxxx}

\vspace{.7cm}

\maketitle

\vspace{1.cm}

\section{Introduction}

\renewcommand{\theequation}{1.\arabic{equation}}
\setcounter{equation}{0}

Brane worlds have been studied extensively in the past decade \cite{branes}, following
Horava and Witten's (HW) ideas \cite{HW96}, where gauge fields of the standard model 
(SM) are confined on two  9-branes located at the end points of an $S^{1}/Z_{2}$ orbifold.
Out of the 9-spatial dimensions of the branes, six are compactified on a very
small scale close to the fundamental one. A 5-dimensional effective theory of the
11-dimensional HW heterotic M-Theory on $S^{1}/Z_{2}$ was worked out explicitly
by Lukas {\em et al} \cite{LOSW99}, and shown that the  radion is stable \cite{LSS05,%
WGW08}, and its mass is of the order of $0.1 \; GeV$ \cite{WGW08}. In addition,  
the corresponding tensor perturbations were also studied, and 
found that the gravity is localized in the visible (TeV) brane \cite{WGW08}. The spectrum  
of the gravitational Kaluza-Klein (KK) towers is discrete, and the  mass gap can be 
in the order of  $TeV$. The corrections to the 4-dimensional Newtonian potential, 
due to the high order KK modes, are exponentially suppressed, and are consistent 
with observations \cite{WGW08}. In such a setup,  the long standing {\em hierarchy problem}, 
namely the large difference in magnitudes between the Planck and electroweak scales,
may  be  potentially resolved by combining the large extra dimension \cite{ADD98},
warped-factor  \cite{RS1} and brane-tension coupling  \cite{Cline99} mechanisms. 
One of the most attractive features of the   model, similar to the RS1 model \cite{RS1},  is that 
it might be soon explored by LHC \cite{DHR00}. For critical reviews of the brane worlds 
and some open issues, we refer readers to \cite{branes}.  

Another important application of brane worlds is to the cosmological constant
problem \cite{wen}. In the 4-dimensional spacetimes, there exists Weinberg's no-go 
theorem for the adjustment of the cosmological constant. However, in higher dimensional 
spacetimes, the 4-dimensional vacuum energy on the brane does not necessarily give 
rise to an effective 4-dimensional cosmological constant. Instead, it may only curve 
the bulk, while leaving the brane still flat \cite{CEG01}, whereby Weinberg's no-go 
theorem is evaded. Along this vein, the cosmological constant problem was studied 
in the framework of brane worlds in 5-dimensional spacetimes \cite{5CC} and 6-dimensional 
supergravity \cite{6CC}. However, it was soon realized that in the 5-dimensional case 
hidden fine-tunings are required \cite{For00}. In the 6-dimensional case such 
fine-tunings may not be needed, but it is still not clear whether loop corrections 
can be as small as  expected \cite{Burg07}.

In addition, by adding an Einstein-Hilbert term to the brane action, Dvali, 
Gabadadze and Porrati (DGP) \cite{DGP} showed that gravity can be altered at immense 
distances, due to the slow leakage of gravity off our 3-dimensional universe into 
bulk. It should be noted that the DGP model has only one 3-brane, and the spacetime
in the direction perpendicular to the brane is usually infinitely large, in contrast 
to the RS1 model, where two orbifold branes form the boundary in the transverse   
direction of the branes, although later Randall and Sundrum proposed another model
(RS2), in which   only one brane exists \cite{RS2}. A remarkable feature of the 
DGP model is that it gives rise to a late cosmic
acceleration of the universe, without the introduction of dark energy \cite{Def}.
It must be noted that, despite of this great success, the DGP model, as well
as its hybrids,  is usually  plagued with the problem  of ghost \cite{ghost1,ghost2}, 
in addition to the problem of the consistency with observations 
\cite{Hu08,GIW09}. 
 
It should also be noted that the RS1, RS2 and DGP brane worlds, as well as  their 
generalizations \cite{branes}, are phenomenological  models, and how to 
implement them into string/M theory is still an open question, 
despite of some important efforts along this direction \cite{Ben99,Chen06}. Such an 
implementation turns out to be extremely difficult, as one would expect, given the 
complexity of the theory. It was exactly because of this that most of the previous 
works on brane worlds are phenomenological, and should be considered only
as an intermediary bridge between observations and fundamental theory. 

Lately, as part of the efforts of implementing the RS1 model into string/M theory,
the orbifold branes  and their applications to cosmology  were studied  systematically
in the framework
of both the Horava-Witten heterotic M-Theory  \cite{GWW07,WGW08} and string
theory \cite{WS07,WSVW08,WS08}  on $S^{1}/Z_{2}$. From the point of view of
pure numerology, it was found that  the   4D  effective cosmological constant  
can be cast in the form,  
\bq
\lb{1.5}
\rho_{\Lambda}
= \frac{\Lambda_{4}}{8\pi G_{4}}
= 3\left(\frac{R}{l_{pl}}\right)^{\alpha_{R}}\left(\frac{M}{M_{pl}}\right)^{\alpha_{M}}
M_{pl}^{4},
\eq
where $R$ denotes the typical size of the extra dimensions, $M$ the energy scale of string or M theory,
 and $(\alpha_{R}, \alpha_{M}) = (10, 16)$ for string
theory \cite{WS07} and $(\alpha_{R}, \alpha_{M}) = (12, 18)$ for the HW heterotic M Theory 
\cite{GWW07}. In both 
cases, it can be shown that for $R \simeq 10^{-22} \; m$ and $M
\simeq 1\; TeV$, we obtain $\rho_{\Lambda} \sim \rho_{\Lambda, ob} \simeq 10^{-47}\; 
GeV^{4}$.  In contrast to that in Einstein's theory, the domination of this term  is only temporary.  
Due to the interaction of the bulk and the brane, the universe will be in its decelerating expansion 
phase again, whereby all problems connected with a far future de Sitter universe \cite{Fish,KS00} 
are resolved. This feature was also found in the DGP model \cite{DGP}. Therefore, a late
transient acceleration of the universe seems to be a generic feature of brane worlds.

It was also showed that  the radion is stable, and its mass is about $10^{-1}$ GeV in 
the Horava-Witten heterotic M-Theory \cite{WGW08} and $10^{-2}$ GeV in the string
theory \cite{WS08}. The gravity is localized on the visible (TeV) brane. The spectrum of the 
gravitational KK towers is discrete with a mass gap that can be in the order of TeV. The high order 
Yukawa corrections to the 4-dimensional effective Newtonian potential are exponentially 
suppressed.

In this paper,  we shall continuously work along the direction of implementing the 
RS1 model \cite{RS1} into string/M theory. In particular, In Sec. II, starting with 
the Neveu-Schwarz/Neveu-Schwarz (NS/NS) sector of type II string, we first consider
the compactification of  the $\left(D+d_{+} + d_{-}\right)$-dimensional spacetime
on two manifolds $M_{d_{+}}$ and $M_{d_{-}}$, where the topologies of $M_{d_{+}}$ 
and $M_{d_{-}}$ are unspecified. This opens the possibility of having the dilaton
and modulus fields non-zero potentials (masses), which is in contrast to the  toroidal 
compactification considered in \cite{WS07,WSVW08,WS08}, in which   these scalar
fields are always massless \cite{LWC00,BW06,TW09}. After reducing the action to an effective
$D$-dimensional one,  we further compactify one of the $(D-1)$ spatial dimensions 
on an $S^{1}/Z_{2}$ orbifold. Lifting it to the original spacetime, they represent
$(D+d_{+} + d_{-} -2)$-dimensional orbiford branes.
 The corresponding gravitational and matter field 
equations both in the bulk and on the branes are derived separately in Sec. III,
while in Sec. IV such developed formulas are applied to cosmology by setting
$D = 5 = d_{+} + d_{-}$. In particular, the generalized Friedmann equations are
given explicitly on the branes. In Sec. V the radion stability and radion mass
are studied, while in Sec. VI, the tensor perturbations are investigated. It is
found that the radion stable, and the gravity is localized on the visible
brane. Both the radion mass and the mass gap of the gravitational KK towers
can be in the order of $TeV$, by properly choosing the free parameters presented
in the model. The high order Yukawa corrections to the 4-dimensional Newtonian 
potential, due to the high order KK modes, is exponentially suppressed, and can be 
negligible.  The paper is ended with Sec. VII, in which we summarize our main 
results and present some remarks to the future work. 

To have this paper as much
independent as possible, for the sake of reader's convenience, some parts might
be repeated from our previous studies of the problems, although we try to limit these to their
minimum.

Before proceeding further, we would like to note that,  to have a late time accelerating 
universe from string/M-Theory, Townsend and Wohlfarth \cite{townsend} invoked a 
time-dependent compactification  
of pure gravity in higher dimensions with hyperbolic internal space to circumvent 
Gibbons' non-go theorem \cite{gibbons}. Their exact solution   exhibits 
a short period of acceleration. The solution is the zero-flux limit of spacelike branes 
\cite{ohta}. If non-zero flux or forms are turned on,  a transient acceleration exists 
for both compact internal hyperbolic and flat spaces \cite{wohlfarth}. Other 
accelerating solutions  by compactifying more complicated time-dependent internal 
spaces can be found  in \cite{string}. 

\section{The Model}
\renewcommand{\theequation}{2.\arabic{equation}}
\setcounter{equation}{0}

In this section, we  consider the  compactification of the NS/NS 
sector  in ($D+d_{+}+d_{-}$)-dimensions, and obtain an effective $D$-dimensional action. 
Then, we  compactify one of the $(D-1)$ spatial dimensions by introducing two orbifold branes 
as the boundaries along this compactified dimension.

\subsection{Compactification of the  NS/NS sector }

Let us consider  the  NS/NS sector  in ($D+d_{+}+d_{-}$)-dimensions, 
$\hat{M}_{N} = M_{D}\times {\cal{M}}_{d_{+}}\times {\cal{M}}_{d_{-}}$, where  
${\cal{M}}_{d_{+}}$ and ${\cal{M}}_{d_{-}}$ are  $d_{+}$ and $d_{-}$ dimensional spaces,
respectively, and $N \equiv D+d_{+}+d_{-}$. To have our formulas as much  applicable as possible, 
we shall not specify the
topologies of these spaces.   The  action takes the form 
\cite{LWC00,BW06,MG07},

\bqn
\lb{2.1}
\hat{S}_{N} &=& - \frac{1}{2\kappa^{2}_{N}}
\int{d^{N}x\sqrt{\left|\hat{g}_{N}\right|}  e^{-\hat{\Phi}}} \nb\\
& & \times \left\{{\hat{R}}_{N}[\hat{g}] +  \left(\hat{\nabla}\hat{\Phi}\right)^{2} 
 - \frac{1}{12}{\hat{H}}^{2}\right\},\;\;\;
\eqn
where $\hat{\nabla}$ denotes the covariant derivative with  respect to $\hat{g}^{AB}$
with  $A, B = 0, 1, ..., N -1$, and $\hat{\Phi}$ is the dilaton field. The NS three-form 
field $\hat{H}_{ABC}$ is defined as
\bqn
\lb{2.1a}
\hat{H}_{ABC} &=& 3 \partial_{[A}\hat{B}_{BC]}\nb\\
&=& \partial_{A}\hat{B}_{BC} + \partial_{B}\hat{B}_{CA}
+ \partial_{C}\hat{B}_{AB},
\eqn
where  the square brackets imply total antisymmetrization over all indices, and
\bq
\lb{2.1b}
 \hat{B}_{CD} = - \hat{B}_{DC}, \;\;\;
 \partial_{A}\hat{B}_{CD}
 \equiv \frac{\partial\hat{B}_{CD}}{\partial{x^{A}}}. 
 \eq
The constant $\kappa^{2}_{N}$ denotes the gravitational coupling constant, defined as
\bq
\lb{2.2}
\kappa^{2}_{N} = 8\pi G_{N} = \frac{1}{M_{N}^{N-2}},
\eq
where $G_{N}$ and $M_{N}$ denote, respectively,  
the $N$-dimensional Newtonian constant and  Planck mass. 

In this paper we consider the $N$-dimensional spacetimes described by the metric,
\bqn
\lb{2.3}
d{\hat{s}}^{2}_{N} &=& \hat{g}_{AB} dx^{A}dx^{B} \nb\\
&=&  \tilde{g}_{ab}\left(x\right) dx^{a}dx^{b} 
 +   e^{\sqrt{\frac{2}{d_{+}}}\psi_{+}(x)}h^{+}_{ij}\left(z_{+}\right)dz^{i}_{+} dz^{j}_{+}\nb\\
  & & +    e^{\sqrt{\frac{2}{d_{-}}}\psi_{-}(x)}h^{-}_{pq}\left(z_{-}\right)dz_{-}^{p} dz_{-}^{q},
\eqn
where  $\tilde{g}_{ab}(x)$ is the metric on $M_{D}$, parametrized by the coordinates $x^{a}$
with  $a,b, c = 0, 1, ..., D-1$,  $h^{+}_{ij}\left(z_{+}\right)$ the metric on the compact space 
${\cal{M}}_{d_{+}}$
with  coordinates $z_{+}^{i}$, where $i, j = D, D+1, ..., D+d_{+}-1$, and $h^{-}_{ij}\left(z_{-}\right)$ 
the metric 
on the compact space ${\cal{M}}_{d_{-}}$ with  coordinates $z_{-}^{p}$, where $p, q = D + d_{+}, D+ d_{+}+1, 
..., N-1$. 

We assume that the daliton field $\hat{\Phi}$ is function of $x^{a}$, and 
the flux $\hat{B}_{CD}$  is block diagonal, 
\bqn
\lb{2.5}
\left( \hat{B}_{CD}\right) &=& \left(\matrix{{B}_{ab}(x) & 0 & 0\cr
0 & e^{\xi_{+}(x)}B_{ij}\left(z_{+}\right) & 0\cr
0 & 0 & e^{\xi_{-}(x)}B_{pq}\left(z_{-}\right)}\right).\nb\\
\eqn
Then, it can be shown that the non-vanishing components of $\hat{H}_{ABC}$ are
\bqn
\lb{2.5b}
\hat{H}_{abc} &=& {H}_{abc} = 3 \partial_{[a}{B}_{bc]}, \nb\\
\hat{H}_{ijk} &=&  e^{\xi_{+}}  H_{ijk} = 3e^{\xi_{+}}  \partial_{[i}{B}_{jk]}, \nb\\
\hat{H}_{pqr} &=&   e^{\xi_{-}}  H_{pqr} = 3 e^{\xi_{-}}  \partial_{[p}{B}_{qr]}, \nb\\
\hat{H}_{aij} &=& B_{ij} e^{\xi_{+}} \tilde{\nabla}_{a}{\xi_{+}}, \nb\\
\hat{H}_{apq} &=& B_{pq} e^{\xi_{-}} \tilde{\nabla}_{a}{\xi_{-}},  
\eqn
where  $\tilde{\nabla}_{a}$ denotes the covariant derivative with  respect to 
$\tilde{g}^{ab}$. On the other hand, we also have   
\bqn
\lb{2.6}
{\hat{R}}_{N}[\hat{g}] &=& \tilde{R}_{D}[\tilde{g}] 
+ e^{-\sqrt{\frac{2}{d_{+}}}\psi_{+}}R_{d_{+}}\left[h^{+}\right]\nb\\
& &
+ e^{-\sqrt{\frac{2}{d_{-}}}\psi_{-}}R_{d_{-}}\left[h^{-}\right]\nb\\
& & - 2\tilde{g}^{ab}\tilde{\nabla}_{a}\tilde{\nabla}_{b}Q 
- \frac{(d_{+}+1)}{2}\left(\tilde{\nabla}\psi_{+}\right)^{2}\nb\\
& &
- \frac{(d_{-}+1)}{2}\left(\tilde{\nabla}\psi_{-}\right)^{2}\nb\\
& &
- \sqrt{d_{+}d_{-}}\left(\tilde{\nabla}\psi_{+}\right)\left(\tilde{\nabla}\psi_{-}\right),
\eqn  
where
\bq
\lb{2.7}
Q \equiv \sqrt{\frac{d_{+}}{2}} \; \psi_{+} + \sqrt{\frac{d_{-}}{2}}\; \psi_{-}.
\eq

Making the following conformal transformations,
\bq
\lb{2.8}
g_{ab} = \Omega^{2}\tilde{g}_{ab},\;\;\;
\Omega = e^{\frac{Q - \hat{\Phi}}{D-2}},
\eq
we find that  
\bqn
\lb{2.9}
\tilde{R}_{D}[\tilde{g}] &=& \Omega^{2}\left\{R_{D}[g] + 2(D-1)\Box\ln\Omega\right.\nb\\
& & \left.
- (D-2)(D-1)\left(\nabla\ln\Omega\right)^{2}\right\},\nb\\
\tilde{g}^{ab}\tilde{\nabla}_{a}\tilde{\nabla}_{b}Q &=& \Omega^{2}\left(\Box{Q}\right.\nb\\
& & \left. 
- (D-2) \left(\nabla{Q}\right)\left(\nabla\ln\Omega\right)\right),
\eqn
where $\Box \equiv g^{ab}\nabla_{a}\nabla_{b}$, and ${\nabla}_{a}$ denotes the covariant 
derivative with  respect to ${g}^{ab}$. Then, 
combining Eqs.(\ref{2.6}) and (\ref{2.9}), we obtain  
\bqn
\lb{2.10}
& & \sqrt{\left|\hat{g}_{N}\right|}  e^{-\hat{\Phi}}  
\left\{{\hat{R}}_{N}[\hat{g}] + \left(\hat{\nabla}\hat{\Phi}\right)^{2}
 - \frac{1}{12}{\hat{H}}^{2}\right\}\nb\\
& & = \sqrt{\left|{g}_{D} {h^{+}}  {h^{-}} \right|}\left\{R_{D}[g]
 + e^{-2\frac{Q - \hat{\Phi}}{D-2}}\left(e^{-\sqrt{\frac{2}{d_{+}}}\psi_{+}}R_{d_{+}}\right.\right.\nb\\
& & \left. + e^{-\sqrt{\frac{2}{d_{-}}}\psi_{-}}R_{d_{-}} - \frac{1}{12}{\hat{H}}^{2}\right)
 + \frac{2}{D-2}\Box{Q}\nb\\
& & - \frac{2(D-1)}{D-2}\Box{\hat{\Phi}}
- \frac{1}{D-2}\left(\nabla{\left(Q - \hat{\Phi} \right)}\right)^{2}\nb\\
& & \left. - \frac{1}{2}\left(\nabla\psi_{+}\right)^{2}  
- \frac{1}{2}\left(\nabla\psi_{-}\right)^{2}\right\},
\eqn
where  
\bqn
\lb{2.11}
\hat{H}^{2} &=& e^{\frac{6\left(Q - \hat{\Phi}\right)}{D-2}}H^{2}\nb\\
& & + 3e^{\frac{2\left(Q - \hat{\Phi}\right)}{D-2}}\left(e^{2\left(\xi_{+} 
- \sqrt{\frac{2}{d_{+}}}\psi_{+}\right)}
B_{+}^{2}\left(\nabla\xi_{+}\right)^{2} \right.\nb\\
& & \left. + e^{2\left(\xi_{-} - \sqrt{\frac{2}{d_{-}}}\psi_{-}\right)}
B_{-}^{2}\left(\nabla\xi_{-}\right)^{2}\right)\nb\\
& & + e^{ 2\xi_{+} - 3\sqrt{\frac{2}{d_{+}}}\psi_{+}}H^{2}_{+} \nb\\
& &
+ e^{ 2\xi_{-} - 3\sqrt{\frac{2}{d_{-}}}\psi_{-}}H^{2}_{-},
\eqn
with
\bqn
\lb{2.12}
H^{2}  &=&    H_{abc}(x) H^{abc}(x),\nb\\ 
H^{2}_{+} &=&   H_{ijk}\left(z_{+}\right) H^{ijk}\left(z_{+}\right),\nb\\
H^{2}_{-} &=&   H_{pqr}\left(z_{-}\right) H^{pqr}\left(z_{-}\right),\nb\\
B^{2}_{+} &=&    B_{ij}\left(z_{+}\right) B^{ij}\left(z_{+}\right),\nb\\
B^{2}_{-} &=&    B_{pq}\left(z_{-}\right) B^{pq}\left(z_{-}\right),
\eqn
and
\bq
\lb{2.12a}
g^{ab} g_{ac} =  \delta^{b}_{c},\;\;\;
h^{+ik} h^{+}_{ij} =  \delta^{k}_{j},\;\;\;
h^{- pq} h^{-}_{pr} =  \delta^{q}_{r}.
\eq
  
Substituting Eqs.(\ref{2.11}) and (\ref{2.12}) into Eq.(\ref{2.1}), and then integrating
it by part, we obtain the $D-$dimensional effective action in the Einstein frame,
\bq
\lb{2.13}
S_{D}^{(E)} = - \frac{1}{2\kappa_{D}^{2}}\int{\sqrt{\left|g_{D}\right|}d^{D}x\left(R_{D}[g]
- {\cal{L}}^{(E)}_{D}\left(\phi_{n}, \xi_{\pm}\right)\right)},
\eq
where $\phi_{n} = \left\{\phi, \psi_{\pm}\right\}$, and  
\bqn
\lb{2.14a}
\kappa_{D}^{2} &\equiv& \frac{\kappa_{N}^{2}}{V_{d_{+}}V_{d_{-}}},\\
\lb{2.14b}
V_{d_{\pm}} &\equiv&\int{\sqrt{\left|h^{\pm} \right|}\; d^{d_{\pm}}z_{\pm}},\nb\\
\lb{2.14c}
{\cal{L}}^{(E)}_{D}  &=&
\frac{1}{2}\sum_{n}{\left(\nabla\phi_{n}\right)^{2}}
+ \frac{1}{12} e^{-\sqrt{\frac{8}{D-2}}\phi}H^{2}\nb\\
& &
+  \alpha_{+}e^{2\xi_{+} - \sqrt{\frac{8}{{d_{+}}}}\psi_{+}}\left(\nabla\xi_{+}\right)^{2}\nb\\
& &
+  \alpha_{-}e^{2\xi_{-} - \sqrt{\frac{8}{{d_{-}}}}\psi_{-}}\left(\nabla\xi_{-}\right)^{2}\nb\\
& & 
- e^{\sqrt{\frac{2}{D-2}}\phi}\left(\beta_{+}e^{- \sqrt{\frac{2}{d_{+}}}\psi_{+}}\right.\nb\\
& &
+ \beta_{-}e^{- \sqrt{\frac{2}{d_{-}}}\psi_{-}}
-    \gamma_{+} e^{2\xi_{+}- \sqrt{\frac{18}{{d_{+}}}}\psi_{+}}  \nb\\
& &  
\left. -     \gamma_{-} e^{2\xi_{-} - \sqrt{\frac{18}{d_{-}}}\psi_{-}}\right),\;\;
\eqn
and  
\bqn
\lb{2.14d}
\phi  &\equiv&  \sqrt{\frac{2}{D-2}}\left(\hat{\Phi} - Q\right),\\
\lb{2.14e}
\alpha_{\pm} &\equiv& \frac{1}{4V_{d_{\pm}}}\int{d^{d_{\pm}}z_{\pm}\; \sqrt{\left|h^{\pm}\right|} 
                B^{2}_{\pm}\left(z_{\pm}\right)},\nb\\
\beta_{\pm} &\equiv& \frac{1}{V_{d_{\pm}}}\int{d^{d_{\pm}}z_{\pm} \; \sqrt{\left|h^{\pm}\right|} 
                R_{d_{\pm}}\left(z_{\pm}\right)},\nb\\
\gamma_{\pm} &\equiv& \frac{1}{12V_{d_{\pm}}}\int{d^{d_{\pm}}z_{\pm}\; \sqrt{\left|h^{\pm}\right|} 
               H^{2}_{\pm}\left(z_{\pm}\right)}.
\eqn

\subsection{$S^{1}/Z_{2}$ Compactification of the D-Dimensional Sector}

We shall compactify  one of the $(D-1)$ spatial dimensions by placing two orbifold branes
as its boundaries. The brane actions are taken as,
\bqn
\lb{3.1}
S^{(E, I)}_{D-1, m} &=& - \epsilon_{I} \int_{M^{(I)}_{D-1}}{\sqrt{\left|g^{(I)}_{D-1}\right|}
V^{(I)}_{D-1}\left(\phi_{n}, \xi_{\pm}\right) d^{D-1}\xi_{(I)}} \nb\\
& & + \int_{M^{(I)}_{D-1}}{d^{D-1}\xi_{(I)}\sqrt{\left|g^{(I)}_{D-1}\right|}}\nb\\
& & \times {\cal{L}}^{(I)}_{D-1,m}\left(\phi_{n}, \xi_{\pm}, \chi\right),
\eqn
where $I = 1, 2,\; V^{(I)}_{D-1}\left(\phi_{n}, \xi_{\pm}\right)$  denotes the potential of the scalar 
fields $\phi_{n}$  on the branes, and $\xi_{(I)}^{\mu}$'s are the intrinsic coordinates of the 
branes with  $\mu, \nu = 0, 1, 2, ..., D-2$, and  $\epsilon_{1} = - \epsilon_{2} = 1$.
$\chi$  denotes collectively the matter fields. The two branes are localized on the surfaces,
\bq
\lb{3.3d}
\Phi_{I}\left(x^{a}\right)  = 0,
\eq 
or equivalently
\bq
\lb{3.3db}
x^{a} = x^{a}\left(\xi^{\mu}_{(I)}\right).
\eq 
$g^{(I)}_{D-1}$ denotes the determinant of the reduced metric  $g_{\mu\nu}^{(I)}$  of the I-th  brane, 
defined as
\bq
\lb{3.3c}
g_{\mu\nu}^{(I)} \equiv \left. g_{ab} e^{(I)a}_{(\mu)} e^{(I)b}_{(\nu)}\right|_{M^{(I)}_{D-1}}, 
\eq
where
\bq
\lb{3.3dd}
e^{(I)\; a}_{(\mu)} \equiv \frac{\partial x^{a}}{\partial \xi^{\mu}_{(I)}}. 
\eq

Then,  the total action is given by,
\bq
\lb{3.2}
S^{(E)}_{total} = S_{D}^{(E)}  + \sum_{I=1}^{2}{S^{(E, I)}_{D-1, m}}.
\eq

\section{Field Equations Both Outside and on the Orbifold Branes}
\renewcommand{\theequation}{3.\arabic{equation}}
\setcounter{equation}{0}

Variation of the total action (\ref{3.2}) with respect to   the metric ${g}_{ab}$ yields 
the  field equations,
\bqn
\lb{3.3}
G^{(D)}_{ab} &=& \kappa^{2}_{D}T^{(D)}_{ab} + \kappa^{2}_{D} 
\sum^{2}_{I=1}{{\cal{T}}^{(I)}_{\mu\nu} e^{(I, \; \mu)}_{a}e^{(I, \; \nu)}_{b}}\nb\\
& &  \times \sqrt{\left|\frac{g^{(I)}_{D-1}}{g_{D}}\right|} \;
\delta\left(\Phi_{I}\right),
\eqn
where  $\delta(x)$ denotes the Dirac delta function, normalized in the sense of \cite{LMW01},
and the energy-momentum tensors $T^{(D)}_{ab}$ and ${\cal{T}}^{(I)}_{\mu\nu}$ are defined as,
\bqn
\lb{3.3a}
\kappa^{2}_{D}T^{(D)}_{ab} &\equiv & \frac{1}{2}
 \left(\nabla_{a}\phi^{n}\right)\left(\nabla_{b}\phi_{n}\right)\nb\\
 & &
 + \alpha_{+}e^{2\xi_{+} - \sqrt{\frac{8}{{d_{+}}}}\psi_{+}}
   \left(\nabla_{a}\xi_{+}\right)\left(\nabla_{b}\xi_{+}\right)\nb\\
 & &
 + \alpha_{-}e^{2\xi_{-} - \sqrt{\frac{8}{d_{-}}}\psi_{-}}
   \left(\nabla_{a}\xi_{-}\right)\left(\nabla_{b}\xi_{-}\right)\nb\\
& & + \frac{1}{4}e^{-\sqrt{\frac{8}{D-2}}\; \phi} H_{acd}H_{b}^{\;\; cd} \nb\\
& &
-  \frac{1}{2}g_{ab} {\cal{L}}^{(E)}_{D},\\
\lb{3.3aa}
{\cal{T}}^{(I)}_{\mu\nu} &\equiv& {\cal{S}}^{(I)}_{\mu\nu} 
+  \tau^{(I)}_{p} g_{\mu\nu}^{(I)},\nb\\
{\cal{S}}^{(I)}_{\mu\nu} &\equiv& 2 \frac{\delta{\cal{L}}^{(I)}_{D-1,m}}
{\delta{g^{(I)\; \mu\nu}}}
-  g^{(I)}_{\mu\nu}{\cal{L}}^{(I)}_{D-1,m},
\eqn
where $\phi^{n} = \phi_{n}$, and 
\bq
\lb{3.3b}
\tau^{(I)}_{p} \equiv  \epsilon_{I} V^{(I)}_{D-1}\left(\phi_{n},\xi_{\pm}\right).
\eq

Variation of the total action (\ref{3.2}), respectively, with  respect to 
$\phi, \; \psi_{\pm}, \; \xi_{\pm}$ and $B_{ab}$, yields the following equations of the 
matter fields,  
\bqn
\lb{3.ea}
\Box\phi &=&  - \frac{1}{12}\sqrt{\frac{8}{D-2}}e^{-\sqrt{\frac{8}{D-2}}\; \phi}H^{2}
 \nb\\
& & -  \sqrt{\frac{2}{D-2}}e^{\sqrt{\frac{8}{D-2}}\; \phi}
\left(\beta_{+}e^{-\sqrt{\frac{2}{d_{+}}}\; \psi_{+}} \right.\nb\\
    & &
     +\beta_{-}e^{-\sqrt{\frac{2}{d_{-}}}\; \psi_{-}} 
     -\gamma_{+}e^{2\xi_{+} -\sqrt{\frac{18}{d_{+}}}\; \psi_{+}}\nb\\
      & & \left.
     -\gamma_{-}e^{2\xi_{-} -\sqrt{\frac{18}{d_{-}}}\; \psi_{-}}\right)\nb\\
     & &
  - \sum^{2}_{i=1}{\left(2\kappa_{D}^{2}\epsilon_{I} \frac{\partial{V_{D-1}^{(I)}}}
{\partial{\phi}} + \sigma^{(I)}_{\phi}\right)}\nb\\
& & \times\sqrt{\left|\frac{g^{(I)}_{D-1}}{g_{D}}\right|} \; \delta\left(\Phi_{I}\right),\\
\lb{3.eb}
\Box\psi_{\pm} &=& - \alpha_{\pm} \sqrt{\frac{8}{d_{\pm}}}
e^{2\xi_{\pm} - \sqrt{\frac{8}{d_{\pm}}}\psi_{\pm}}\left(\nabla\xi_{\pm}\right)^{2}\nb\\
& & e^{\sqrt{\frac{2}{D-2}}\;\phi}\left(\beta_{\pm}\sqrt{\frac{2}{d_{\pm}}}
    e^{- \sqrt{\frac{2}{d_{\pm}}}\psi_{\pm}}\right.\nb\\
 & & \left.
    - \gamma_{\pm} \sqrt{\frac{18}{d_{\pm}}}  
    e^{2\xi_{\pm} - \sqrt{\frac{18}{d_{\pm}}}\psi_{\pm}}\right)\nb\\
 & &   - \sum^{2}_{i=1}{\left(2\kappa_{D}^{2}\epsilon_{I} \frac{\partial{V_{D-1}^{(I)}}}
{\partial{\psi_{\pm}}} + \sigma^{(I)}_{\psi_{\pm}}\right)}\nb\\
& & \times\sqrt{\left|\frac{g^{(I)}_{D-1}}{g_{D}}\right|} \; \delta\left(\Phi_{I}\right),\\
\lb{3.ec}
\Box \xi_{\pm} &=&  - \left(\nabla\xi_{\pm}\right)^{2} +
\sqrt{\frac{8}{d_{\pm}}} \left(\nabla_{a}\xi_{\pm}\right)\left(\nabla^{a}\psi_{\pm}\right)\nb\\
& & + \frac{\gamma_{\pm}}{\alpha_{\pm}}e^{\sqrt{\frac{2}{D-2}}\;\phi 
    - \sqrt{\frac{2}{d_{\pm}}}\psi_{\pm}} - \frac{1}{2\alpha_{\pm}}
    e^{\sqrt{\frac{8}{d_{\pm}}}\psi_{\pm} - 2\xi_{\pm}}\nb\\
& & \times \sum^{2}_{I=1}{\left(2\kappa_{D}^{2}\epsilon_{I} \frac{\partial{V_{D-1}^{(I)}}}
{\partial{\xi_{\pm}}} + \sigma^{(I)}_{\xi_{\pm}}\right)}\nb\\
& & \times
\sqrt{\left|\frac{g^{(I)}_{D-1}}{g_{D}}\right|} \; \delta\left(\Phi_{I}\right),\\
\lb{3.ed}
\nabla^{c}H_{cab} &=& \sqrt{\frac{8}{D-2}}\; H_{cab}\nabla^{c}\phi\nb\\
& & -  \sum^{2}_{i=1}{\sigma^{(I)}_{ab}
\sqrt{\left|\frac{g^{(I)}_{D-1}}{g_{D}}\right|} \; \delta\left(\Phi_{I}\right)},
\eqn
where 
\bqn
\lb{3.ee}
\sigma_{\phi}^{(I)} &\equiv& - 2\kappa^{2}_{D}\frac{\delta{\cal{L}}^{(I)}_{D-1,m}}{\delta\phi},\nb\\
\sigma_{\psi_{\pm}}^{(I)} &\equiv& - 2\kappa^{2}_{D}
           \frac{\delta{\cal{L}}^{(I)}_{D-1,m}}{\delta\psi_{\pm}},\nb\\
\sigma_{\xi_{\pm}}^{(I)} &\equiv& - 2\kappa^{2}_{D}
           \frac{\delta{\cal{L}}^{(I)}_{D-1,m}}{\delta\xi_{\pm}},\nb\\
\sigma^{(I)}_{ab} &\equiv& - 4\kappa^{2}_{D} e^{\sqrt{\frac{8}{D-2}}\; \phi}
\frac{\delta{\cal{L}}^{(I)}_{D-1,m}}{\delta{B^{ab}}}.
\eqn
 
Eq.(\ref{3.3}) and Eqs.(\ref{3.ea})-(\ref{3.ed}) consist of the complete set of the gravitational 
and matter field equations.
To solve these equations, it is found very convenient to separate them  into two groups, one is 
defined outside the two orbifold branes, and the other is defined on the two branes. 

\subsection{Field Equations Outside the Two Branes}

To write down the equations outside the two orbifold branes is straightforward, and they are simply  the 
D-dimensional gravitational field equations (\ref{3.3}), and the matter field equations
Eqs.(\ref{3.ea})-(\ref{3.ed})  without the delta function parts,  
\bqn
\lb{3.ef}
\Box\phi &=&  - \frac{1}{12}\sqrt{\frac{8}{D-2}}e^{-\sqrt{\frac{8}{D-2}}\; \phi}H^{2}
 \nb\\
& & -  \sqrt{\frac{2}{D-2}}e^{\sqrt{\frac{8}{D-2}}\; \phi}
\left(\beta_{+}e^{-\sqrt{\frac{2}{d_{+}}}\; \psi_{+}} \right.\nb\\
    & &
     +\beta_{-}e^{-\sqrt{\frac{2}{d_{-}}}\; \psi_{-}} 
      -\gamma_{+}e^{2\xi_{+} -\sqrt{\frac{18}{d_{+}}}\; \psi_{+}} \nb\\
      & & \left.
     -\gamma_{-}e^{2\xi_{-} -\sqrt{\frac{18}{d_{-}}}\; \psi_{-}}\right),\\  
\lb{3.eg}
\Box\psi_{\pm} &=& - \alpha_{\pm} \sqrt{\frac{8}{d_{\pm}}}
e^{2\xi_{\pm} - \sqrt{\frac{8}{d_{\pm}}}\psi_{\pm}}\left(\nabla\xi_{\pm}\right)^{2}\nb\\
& & e^{\sqrt{\frac{2}{D-2}}\;\phi}\left(\beta_{\pm}\sqrt{\frac{2}{d_{\pm}}}
    e^{- \sqrt{\frac{2}{d_{\pm}}}\psi_{\pm}}\right.\nb\\
 & & \left.
    - \gamma_{\pm} \sqrt{\frac{18}{d_{\pm}}}  
    e^{2\xi_{\pm} - \sqrt{\frac{18}{d_{\pm}}}\psi_{\pm}}\right),\\
\lb{3.eh}
\Box \xi_{\pm} &=&  - \left(\nabla\xi_{\pm}\right)^{2} +
\sqrt{\frac{8}{d_{\pm}}} \left(\nabla_{a}\xi_{\pm}\right)\left(\nabla^{a}\psi_{\pm}\right)\nb\\
& & + \frac{\gamma_{\pm}}{\alpha_{\pm}}e^{\sqrt{\frac{2}{D-2}}\;\phi 
    - \sqrt{\frac{2}{d_{\pm}}}\psi_{\pm}},\\
\lb{3.ei}
\nabla^{c}H_{cab} &=& \sqrt{\frac{8}{D-2}}\; H_{cab}\nabla^{c}\phi.
\eqn
Therefore, in the rest of this section, we shall 
concentrate ourselves on the derivation of the field equations on the branes. 

\subsection{Field Equations on the Two Branes}

To write down the field  equations on the two orbifold branes, one can follow two different 
approaches: (i)  First express the delta function parts in the left-hand sides of
Eqs.(\ref{3.3}) and (\ref{3.ea})-(\ref{3.ed}) in terms of the discontinuities of the first 
derivatives of the metric coefficients and matter fields, and then equal the corresponding
delta function parts  in the right-hand sides of these equations, as shown systematically in 
\cite{Bin01,WCS06}. (ii) The second approach is to use the Gauss-Codacci and Lanczos equations 
to write down the $(D-1)$-dimensional gravitational field equations on the branes  \cite{SMS}.
 It should be noted 
that these two  approaches are  equivalent and complementary one to the other. In this paper, 
we shall follow the second approach to write down the gravitational field equations on the 
two branes, and the first approach to write the matter field equations on the two branes.

\subsubsection{Gravitational Field Equations on the Two Branes}

From the Gauss-Codacci equations, we we obtain \cite{SMS}, 
 \bq
 \lb{3.12}
 G^{(D-1)}_{\mu\nu} = {\cal{G}}^{(D)}_{\mu\nu} + E^{(D)}_{\mu\nu}
 + {\cal{F}}^{(D-1)}_{\mu\nu},
 \eq
 with 
 \bqn
 \lb{3.13}
 {\cal{G}}_{\mu\nu}^{(D)} &\equiv&  \frac{D-3}{(D-2)}
\left\{G_{ab}^{(D)}e^{a}_{(\mu)} e^{b}_{(\nu)} \right.\nb\\
& & \left.
- \left[G_{ab}n^{a}n^{b} + \frac{1}{D-1} G^{(D)}\right]g_{\mu\nu}\right\}, \nb\\
E^{(D)}_{\mu\nu} &\equiv& C_{abcd}^{(D)}n^{a}e^{b}_{(\mu)}n^{c}e^{d}_{(\nu)},\nb\\
{\cal{F}}^{(D-1)}_{\mu\nu} &\equiv&  
 K_{\mu\lambda}K^{\lambda}_{\nu} - KK_{\mu\nu} \nb\\
& & - \frac{1}{2}g_{\mu\nu}\left(K_{\alpha\beta}K^{\alpha\beta} 
    - K^{2}\right),
\eqn
where $n^{a}$ denotes the normal vector to the brane, $G^{(D)}
\equiv g^{ ab} G^{(D)}_{ab}$, and $C_{abcd}^{(D)}$ the Weyl tensor. 
The extrinsic curvature $K_{\mu\nu}$ is defined as
\bq
\lb{3.13a}
K_{\mu\nu} \equiv e^{a}_{(\mu)}e^{b}_{(\nu)}\nabla_{a}n_{b}.
\eq
A crucial step of this approach is the Lanczos equations \cite{Lan22},
\bq
\lb{3.4}
\left[K_{\mu\nu}^{(I)}\right]^{-} - g_{\mu\nu}^{(I)} \left[K^{(I)}\right]^{-} 
= - \kappa^{2}_{D}{\cal{T}}_{\mu\nu} ^{(I)},
\eq
where 
\bqn
\lb{3.5}
\left[K_{\mu\nu}^{(I)}\right]^{-} &\equiv& {\rm lim}_{\Phi_{I} \rightarrow 0^{+}}
K^{(I)\; +}_{\mu\nu} - {\rm lim}_{\Phi_{I} \rightarrow 0^{-}}
K^{(I)\; -}_{\mu\nu},\nb\\
\left[K^{(I)}\right]^{-} &\equiv& g^{(I)\; \mu\nu}\left[K_{\mu\nu}^{(I)}\right]^{-}.
\eqn

Assuming that the branes have $Z_{2}$ symmetry, we can express the intrinsic
curvatures $K^{(I)}_{\mu\nu}$ in terms of the effective energy-momentum tensor
${\cal{T}}_{\mu\nu} ^{(I)}$ through the Lanczos equations (\ref{3.4}). Setting
\bq
\lb{3.14}
 {\cal{S}}^{(I)}_{\mu\nu} = \tau^{(I)}_{\mu\nu} + g_{k}^{(I)}g^{(I)}_{\mu\nu},
 \eq
 where $g_{k}^{(I)}$ is a coupling constant of the I-th  brane \cite{Cline99}, 
 we find that
 \bq
\lb{3.14aa}
 {\cal{T}}^{(I)}_{\mu\nu} = \tau^{(I)}_{\mu\nu} + \left(g_{k}^{(I)}
 + \tau^{(I)}_{p}\right)g^{(I)}_{\mu\nu}.
 \eq
 Then,
 $ G^{(D-1)}_{\mu\nu}$ given by Eq.(\ref{3.12}) can be cast in the form,
 \bqn
 \lb{3.15}
 G^{(D-1)}_{\mu\nu} &=& {\cal{G}}^{(D)}_{\mu\nu} + E^{(D)}_{\mu\nu}
 + {\cal{E}}_{\mu\nu}^{(D-1)} + \kappa^{4}_{D}\pi_{\mu\nu}\nb\\
 & & + \kappa^{2}_{D-1}\tau_{\mu\nu} + \Lambda_{D-1} g_{\mu\nu},
 \eqn
 where
 \bqn
\lb{3.16}
\pi_{\mu\nu} &\equiv& \frac{1}{4}\left\{\tau_{\mu\lambda}\tau^{\lambda}_{\nu}
-  \frac{1}{D-2}\tau \tau_{\mu\nu}\right.\nb\\
& & \left. 
 - \frac{1}{2}g_{\mu\nu}\left(\tau^{\alpha\beta} \tau_{\alpha\beta}
 - \frac{1}{D-2}\tau^{2}\right)\right\},\nb\\
 {\cal{E}}_{\mu\nu}^{(D-1)} &\equiv& \frac{\kappa^{4}_{D}(D-3)}{4(D-2)}\tau_{p}\nb\\
 & & \times \left[\tau_{\mu\nu}  
   + \left(g_{k} + \frac{1}{2}\tau_{p}\right)g_{\mu\nu}\right],
\eqn
and 
\bqn
\lb{3.17}
\kappa^{2}_{D-1} &=& \frac{D-3}{4(D-2)}g_{k}\kappa^{4}_{D},\nb\\
\Lambda_{D-1}  &=&  \frac{D-3}{8(D-2)}g_{k}^{2}\kappa^{4}_{D}.
\eqn
 For a perfect fluid,
\bq
\lb{3.18}
\tau_{\mu\nu} = \left(\rho + p\right)u_{\mu}u_{\nu} - p g_{\mu\nu},
\eq
where $u_{\mu}$ is the four-velocity of the fluid, we find that 
\bqn
\lb{3.19}
\pi_{\mu\nu} &=&  \frac{D-3}{4(D-2)}\rho\nb\\
& & \times \left[\left(\rho + p\right)u_{\mu}u_{\nu} 
- \left(p + \frac{1}{2}\rho\right)g_{\mu\nu}\right].
\eqn
Note that in writing Eqs.(\ref{3.15})-(\ref{3.19}), without causing any confusion, we had 
dropped the super indices $(I)$.  

\subsubsection{Matter Field Equations on the Two Branes}

On the other hand, the I-th brane, localized on the surface $\Phi_{I}(x) = 0$, divides the
spacetime into two regions, one with $\Phi_{I}(x) > 0$ and the other with $\Phi_{I}(x) < 0$. 
Since the field equations are the second-order differential equations, the
matter fields have to be at least continuous across this surface, although in general their 
first-order directives are not. Introducing the Heaviside function, defined as
\bq
\lb{3.21a}
H\left(x\right) = \left\{\matrix{1, & x > 0,\cr
0, & x < 0,\cr} \right.
\eq
in the neighborhood of $\Phi_{I}(x)  = 0$ we can write the matter fields in the form, 
\bq
\lb{3.20}
F(x) = F^{+}(x) H\left(\Phi_{I}\right) 
       + F^{-}(x)\left[1 - H\left(\Phi_{I}\right)\right],
\eq
where $F \equiv \left\{\phi, \; \psi_{\pm}, \; \xi_{\pm},\; B\right\}$,
and $F^{+}\; (F^{-})$ is defined in the region $\Phi_{I} > 0\; (\Phi_{I} < 0)$. Then, we find
that
\bqn
\lb{3.21}
F_{,a}(x) &=& F^{+}_{,a}(x) H\left(\Phi_{I}\right) 
       + F^{-}_{,a}(x)\left[1 - H\left(\Phi_{I}\right)\right],\nb\\
F_{,ab}(x) &=& F^{+}_{,ab}(x) H\left(\Phi_{I}\right) 
       + F^{-}_{,ab}(x)\left[1 - H\left(\Phi_{I}\right)\right]\nb\\
       & & + \left[F_{,a}\right]^{-}\frac{\partial \Phi_{I}(x)}{\partial x^{b}}\; 
\delta\left(\Phi_{I}\right),
\eqn    
where $\left[F_{,a}\right]^{-}$ is defined as that  in Eq.(\ref{3.5}). Projecting
$F_{,a}$ onto $n^{a}$ and $e^{a}_{(\mu)}$ directions, we find
\bq
\lb{3.22}
F_{,a} = F_{,\mu} e_{a}^{(\mu)} - F_{,n} n_{a},
\eq
where 
\bq
\lb{3.23}
F_{,n} \equiv n^{a}F_{,a},\;\;
F_{,\mu} \equiv e^{a}_{(\mu)}F_{,a}.
\eq
Then, we have
\bqn
\lb{3.24}
& & \left[F_{,a}\right]^{-} n^{a} = \left[F_{,n}\right]^{-},\nb\\
& & \left[F_{,a}\right]^{-} e^{a}_{(\mu)} = 0.
\eqn
Inserting Eqs.(\ref{3.22})-(\ref{3.24}) into Eq.(\ref{3.21}), we find
\bqn
\lb{3.25}
F_{,ab}(x) &=& F^{+}_{,ab}(x) H\left(\Phi_{I}\right) 
       + F^{-}_{,ab}(x)\left[1 - H\left(\Phi_{I}\right)\right]\nb\\
       & & - \left[F_{,n}\right]^{-}n_{a}n_{b} N_{I} 
       \;  \delta\left(\Phi_{I}\right),
\eqn
where $N_{I} \equiv \sqrt{\left|\Phi_{I,c}\Phi_{I}^{,c}\right|}$, and
\bq
\lb{3.26}
n_{a} = \frac{1}{N_{I}}
\frac{\partial \Phi_{I}(x)}{\partial x^{a}}.
\eq
Substituting Eq.(\ref{3.25}) into Eqs.(\ref{3.ea})-(\ref{3.ed}), we find that the matter
field equations on the branes read,  
\bqn
\lb{3.27a}
\left[\phi^{(I)}_{,n}\right]^{-} &=& -  \Psi^{(I)} 
           \left(2\kappa_{D}^{2}\epsilon_{I} \frac{\partial{V_{D-1}^{(I)}}}
{\partial{\phi}} + \sigma^{(I)}_{\phi}\right),\\
\lb{3.27b}
\left[\psi^{(I)}_{\pm,n}\right]^{-} &=& -  \Psi^{(I)} 
              \left(2\kappa_{D}^{2}\epsilon_{I} \frac{\partial{V_{D-1}^{(I)}}}
{\partial{\psi_{\pm}}} + \sigma^{(I)}_{\psi_{\pm}}\right),\\
\lb{3.27c}
\left[\xi^{(I)}_{\pm,n}\right]^{-} &=& - \frac{\Psi^{(I)}}{2\alpha_{\pm}}
    e^{\sqrt{\frac{8}{d_{\pm}}}\psi_{\pm} - 2\xi_{\pm}}\nb\\
& & \times \sum^{2}_{I=1}{\left(2\kappa_{D}^{2}\epsilon_{I} \frac{\partial{V_{D-1}^{(I)}}}
{\partial{\xi_{\pm}}} + \sigma^{(I)}_{\xi_{\pm}}\right)},\\
\lb{3.27d}
\left[H^{(I)}_{nab}\right]^{-} &=& -   \Psi^{(I)}\; \sigma^{(I)}_{ab},
\eqn
where
\bq
\lb{3.28}
H_{nab} \equiv H_{cab} n^{c},\;\;
\Psi^{(I)} \equiv \frac{1}{N_{I}} \sqrt{\left|\frac{g^{(I)}_{D-1}}{g_{D}}\right|}.
\eq

 
This completes our general description for $(D+d_{+} + d_{-})$-dimensional spacetimes of string theory
with  two orbifold branes.

\section{10-Dimensional Spacetimes and Brane Cosmology}
\renewcommand{\theequation}{4.\arabic{equation}}
\setcounter{equation}{0}

In this section, we restrict ourselves to the  10-dimensional spacetimes 
of string theory with $D=5$ and $d_{+} + d_{-} = 5$.  It can be shown that
%
%
the general metric for the five-dimensional spacetime with a  3-dimensional 
spatial space that is homogeneous, isotropic, and independent of time must take the
form \cite{WGW08},
\bq
\lb{4.2}
ds^{2}_{5} = g_{ab} dx^{a}dx^{b}  
 =g_{MN} dx^{M}dx^{N} - e^{2\omega\left(x^{M}\right)}d\Sigma^{2}_{k},
\eq
where  $ M, \; N = 0, 1$.   Choosing the conformal gauge,   
\bq
\lb{4.3a}
g_{00} = g_{11}, \;\;\; g_{01} = 0, 
\eq
we find that   the five-dimensional metric finally takes the form, 
\bq
\lb{4.4}
ds^{2}_{5} =  e^{2\sigma(t,y)}\left(dt^{2} - dy^{2}\right)  
- e^{2\omega(t,y)}d\Sigma^{2}_{k}.
\eq 
It should be noted that metric (\ref{4.4}) is still subjected to the gauge freedom,
\bq
\lb{4.5}
t = f(t' + y') + g(t' - y'), \;\;\; y = f(t' + y') - g(t' - y'),
\eq
where $f(t' + y')$ and $g(t' - y')$ are arbitrary functions of their indicated
arguments. 
 
It should be noted that in \cite{Martin04} comoving branes were considered, and it was claimed that
the gauge freedom of Eq.(\ref{4.5}) can always bring the two branes at rest (comoving). However, 
this excludes colliding branes \cite{TW09, WGW08,Chen06}. In this 
paper, we shall leave this possibility open.

\subsection{Field Equations Outside  the Two Branes}

To have the problem tractable, in the rest of this paper, we  shall  turn off the flux, i.e., 
\bq
\lb{4.6aa}
\hat{B}_{CD} = 0, 
\eq
so that 
\bq
\lb{4.6ab}
\xi_{\pm} = 0, \;\;\;
\alpha_{\pm} = 0, \;\;
\gamma_{\pm} = 0.  
\eq
Then, it can be shown that outside  the two branes
the field equations (\ref{3.3}) have four independent components, 
which can be cast in the form,  
\bqn
\lb{4.6a}
& & \omega_{,tt} + \omega_{,t}\left(\omega_{,t} - 2\sigma_{,t}\right) 
    + \omega_{,yy} + \omega_{,y}\left(\omega_{,y} - 2\sigma_{,y}\right)   \nb\\
        && = -\frac{1}{6}\left(\phi_{,t}^{2} + \phi_{,y}^{2} + \psi_{+,t}^{2} 
	+ \psi_{+,y}^{2} + \psi_{-,t}^{2} + \psi_{-,y}^{2} \right),\\
\lb{4.6b}
& & 2\sigma_{,tt} + \omega_{,tt} - 3 {\omega_{,t}}^{2}
- \left(2\sigma_{,yy} + \omega_{,yy} - 3 {\omega_{,y}}^{2}\right) - 4ke^{2(\sigma-\omega)}  \nb\\
       && =-\frac{1}{2}\left(\phi_{,t}^{2} - \phi_{,y}^{2} + \psi_{+,t}^{2} - \psi_{+,y}^{2} 
       + \psi_{-,t}^{2} - \psi_{-,y}^{2}  \right),\\
\lb{4.6c}
& &  \omega_{,ty} + \omega_{,t} \omega_{,y} 
    -  \left(\sigma_{,t}\omega_{,y} + \sigma_{,y}\omega_{,t}\right) \nb\\
        &&= -\frac{1}{6}\left(\phi_{,t}\phi_{,y} + \psi_{+,t}\psi_{+,y} 
	+  \psi_{-,t}\psi_{-,y} \right),\\
\lb{4.6d}
& & \omega_{,tt} + 3{\omega_{,t}}^{2} - \left(\omega_{,yy} + 3{\omega_{,y}}^{2}\right)
     + 2ke^{2(\sigma-\omega)} \nb\\ 
        &&= \frac{1}{3}e^{2\sigma}V_{5}.
\eqn
 
On the other hand, the Klein-Gordon equations (\ref{3.ef}) and (\ref{3.eg}) take the form,
\bqn
\lb{4.8a}
& & \phi_{,tt} + 3\phi_{,t}\omega_{,t} - \left(\phi_{,yy} + 3\phi_{,y}\omega_{,y}\right)\nb\\
& & \;\;\;\;\;\;\;\;\;\;\;\;\;\;\;   =  -\sqrt{\frac{2}{3}}e^{2\sigma}V_{5},\\
\lb{4.8b}
& & \psi_{+,tt} + 3\psi_{+,t}\omega_{,t} - \left(\psi_{+,yy} + 3\psi_{,y}\omega_{,y}\right)\nb\\
& & \;\;\;\;\;\;\;\;\;\;\;\;\;\;\;   =  e^{2\sigma}V_{5},
\lb{4.8c}\\
& & \psi_{-,tt} + 3\psi_{-,t}\omega_{,t} - \left(\psi_{-,yy} + 3\psi_{,y}\omega_{,y}\right)\nb\\
& & \;\;\;\;\;\;\;\;\;\;\;\;\;\;\;   =  \sqrt{\frac{2}{3}}e^{2\sigma}V_{5},
\eqn
with 
\bq
\lb{4.8d} 
 V_{5} = e^{\sqrt{\frac{2}{3}}\phi}\left( \beta_{+}e^{-\psi_{+}} 
 + \beta_{-}e^{-\sqrt{\frac{2}{3}}\psi_{-}}\right).
\eq

\subsection{Field Equations on the Two Branes}

Eqs.(\ref{4.6a}) - (\ref{4.8c}) are the field equations that are valid in between the two orbifold branes,
$y_{2}(t_{2}) < y < y_{1}(t_{1})$, where $y = y_{I}(t_{I})$ denote the locations of
the two branes. The proper distance between the two branes is given by
\bq
\lb{4.8ca}
{\cal{D}}(t) = \int_{y_{2}}^{y_{1}}{e^{\sigma(t, y)}dy}.
\eq

On each of the two branes, the metric reduces to
\bq
\lb{4.9}
\left. ds^{2}_{5}\right|_{M^{(I)}_{4}} = g^{(I)}_{\mu\nu}d\xi_{(I)}^{\mu}d\xi_{(I)}^{\nu}
= d\tau_{I}^{2} - a^{2}\left(\tau_{I}\right)d\Sigma^{2}_{k},
\eq
where $\xi^{\mu}_{(I)} \equiv \left\{\tau_{I}, r, \theta, \varphi\right\}$, and
$\tau_{I}$ denotes the proper time of the I-th  brane, defined by
\bqn
\lb{4.10}
d\tau_{I} &=& e^{\sigma\left[t_{I}(\tau_{I}), y_{I}(\tau_{I})\right]}
\sqrt{1 - \left(\frac{\dot{y}_{I}}{\dot{t}_{I}}\right)^{2}}\; dt_{I},\nb\\
a\left(\tau_{I}\right) &\equiv& e^{\omega\left[t_{I}(\tau_{I}), y_{I}(\tau_{I})\right]},
\eqn
with  $\dot{y}_{I} \equiv d{y}_{I}/d\tau_{I}$, etc. For the sake of simplicity and without of causing any
confusion, from now on we shall drop all the indices ``I", unless some specific attention is 
needed. Then, the normal vector $n_{a}$ and the tangential vectors $e^{a}_{(\mu)}$ are
given, respectively, by  
\bqn
\lb{4.11}
n_{a} &=&   e^{2\sigma}\left(- \dot{y}\delta^{t}_{a} +  \dot{t}\delta^{y}_{a}\right),\nb\\
n^{a} &=& -   \left(\dot{y}\delta^{a}_{t} +  \dot{t}\delta^{a}_{y}\right),\nb\\
e^{a}_{(\tau)} &=&  \dot{t}\delta^{a}_{t} +  \dot{y}\delta^{a}_{y},
\;\;\; e^{a}_{(r)} = \delta^{a}_{r},\nb\\ 
e^{a}_{(\theta)} &=& \delta^{a}_{\theta},\;\;\;
e^{a}_{(\varphi)} = \delta^{a}_{\varphi}.
\eqn
Then, it can be shown that   
\bqn
\lb{4.12}
{\cal{G}}^{(5)}_{\mu\nu} &=& {\cal{G}}^{(5)}_{\tau} \delta^{\tau}_{\mu} \delta^{\tau}_{\nu} 
 - {\cal{G}}^{(5)}_{\theta} \delta^{m}_{\mu} \delta^{n}_{\nu}g_{mn},\nb\\
 E^{(5)}_{\mu\nu} &=& E^{(5)}\left(3 \delta^{\tau}_{\mu} \delta^{\tau}_{\nu} 
 -  \delta^{m}_{\mu} \delta^{n}_{\nu}g_{mn}\right),
 \eqn
 where   
 \bqn
 \lb{4.13}
 {\cal{G}}^{(5)}_{\tau} &\equiv& \frac{1}{3}e^{-2\sigma}
 \left(\phi_{,t}^2 - \phi_{,y}^2 + \psi_{+,t}^2 - \psi_{+,y}^2 
 + \psi_{-,t}^2 - \psi_{-,y}^2 \right)\nb\\
 & &  -\frac{5}{24}\left[(\nabla\phi)^2 + (\nabla\psi_{+})^2 
 + (\nabla\psi_{-})^2 \right] + \frac{1}{4}V_5 ,\nb\\
 {\cal{G}}^{(5)}_{\theta} &\equiv& \frac{1}{3}\left[\phi_{,n}^2 
 + \psi_{+,n}^2 + \psi_{-,n}^2 \right] \nb\\
 & & + \frac{5}{24}\left[ (\nabla\phi)^2 + (\nabla\psi_{+})^2 
 + (\nabla\psi_{-})^2 \right] - \frac{1}{4}V_5  ,\nb\\
 E^{(5)} &\equiv& \frac{1}{6}e^{-2\sigma}\left[\left(\sigma_{,tt} - \omega_{,tt}\right)
 - \left(\sigma_{,yy} - \omega_{,yy}\right) \right.\nb\\
 & & \left. + k e^{2(\sigma - \omega)}\right],  
 \eqn
with  $\phi_{,n} \equiv n^{a}\nabla_{a}\phi$.  Then, it can be shown that the four-dimensional 
field equations on each of the two branes
take the form,  
\bqn
\lb{4.14a}
 H^{2} &+& \frac{k}{a^{2}} = \frac{8\pi G}{3}\left(\rho + \tau_{p}\right) 
     + \frac{1}{3}\Lambda 
      + \frac{1}{3}{\cal{G}}^{(5)}_{\tau} + E^{(5)}\nb\\
       & &  
       + \frac{2\pi G}{3\rho_{\Lambda}}\left(\rho + \tau_{p}\right)^{2}, \\
\lb{4.14b}
\frac{\ddot{a}}{a}   &=&  - \frac{4\pi G}{3}\left(\rho +3p - 2 \tau_{p}\right) 
      + \frac{1}{3}\Lambda  - E^{(5)}
      \nb\\
      & &  - \frac{1}{6}\left({\cal{G}}^{(5)}_{\tau} + 3{\cal{G}}^{(5)}_{\theta}\right)
      - \frac{2\pi G}{3\rho_{\Lambda}}\left[\rho\left(2\rho + 3p\right) \right.\nb\\
      & &  
      \left. + \left(\rho + 3p
       - \tau_{p}\right) \tau_{p}\right],  
\eqn 
where $H \equiv \dot{a}/{a}, , \; \Lambda \equiv \Lambda_{4}$
and $G \equiv G_{4}$.

On the other hand, from 
Eqs.(\ref{3.27a}) and (\ref{3.27b}), we find that
\bqn
\lb{4.14d}
\left[\phi^{(I)}_{,n}\right]^{-} &=& -    \left(2\kappa_{5}^{2}\epsilon_{I} \frac{\partial{V^{(I)}_{4}}}
{\partial{\phi}} + \sigma^{(I)}_{\phi}\right)\; \Psi,\\
\lb{4.14e}
\left[\psi^{(I)}_{\pm,n}\right]^{-} &=& -    \left(2\kappa_{5}^{2}\epsilon_{I} \frac{\partial{V^{(I)}_{4}}}
{\partial{\psi_{\pm}}} + \sigma^{(I)}_{\psi_{\pm}}\right)\; \Psi. 
\eqn

\section{ Radion stability and radion mass}
\renewcommand{\theequation}{5.\arabic{equation}}
\setcounter{equation}{0}
 
\begin{figure}
\includegraphics[width=\columnwidth]{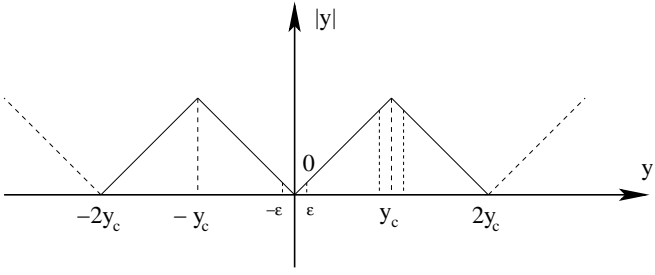}
\caption{The function $\left|y\right|$  appearing in the metric Eq.(\ref{5.1b}).}
\label{fig2}
\end{figure}

In the studies of   branes, an important issue is the  radion stability.
In this section, we shall address this problem. For such a purpose,  let us consider
the 5-dimensional static metric with a 4-dimensional Poincar\'e symmetry, which is
given by Eq.(\ref{4.4}) with $k = 0$ and $\sigma(y) = \omega (y)$, that is, 
\bq
\lb{5.1a}
ds^{2}_{5} = e^{2\sigma(y)}\left(\eta_{\mu\nu}dx^{\mu}dx^{\nu} - dy^{2}\right).
\eq
Then, we find that the corresponding solutions are given by,
\bqn
\lb{5.1b}
\sigma(y) &=&  \frac{1}{3}ln\left(\frac{|y| + y_{0}}{L}\right), \nb\\
\phi(y) &=& c_1ln\left(\frac{|y| + y_{0}}{L}\right) 
           + \phi_0 ,\nb\\
\psi_{+}(y) &=& c_2ln\left(\frac{|y| + y_{0}}{L}\right) + \psi_{+}^0,\nb\\
\psi_{-}(y) &=& \sqrt{\frac{3}{2}}c_2\ln\left(\frac{|y| + y_{0}}{L}\right)  
            + \psi_{-}^0, 
\eqn
where $c_{1}, y_{0},\; L,\; \; \sigma_{0},\; \phi_0$, and $\psi_{+}^0$ are all arbitrary 
constants, and
\bqn
c_{2} &=& \pm \sqrt{\frac{2\left(8-3c_{1}^{2}\right)}{15}},\nb\\
\psi_{-}^0 &=& \sqrt{\frac{3}{2}}\left(\psi_0 - 
\ln\left(\frac{-\beta_{+}}{\beta_{-}}\right)\right).
\eqn
The function $|y|$ is defined as in Fig. \ref{fig2}.

Then, it can be shown that  the above solution satisfies the  gravitational and matter 
field equations outside the branes, Eqs.(\ref{4.6a})-(\ref{4.8b}). On the other hand, to 
show that it also satisfies the field equations on the branes, given by 
Eqs.(\ref{4.14a})-(\ref{4.14b}) and Eqs.(\ref{4.14d})-(\ref{4.14e}), we first note that
the normal vector $n^{a}_{(I)}$  to the I-th brane is given  by
\bq
\lb{5.1d}
n^{a}_{(I)} = - \epsilon_{y}^{(I)} e^{-\sigma(y_{I})}\delta^{a}_{y},
\eq
and that  
\bqn
\lb{5.1e}
\dot{t} &=& e^{-\sigma(y_{I})}, 
\;\; \dot{y} = 0,\nb\\
{\cal{G}}_{\tau}^{(5)} &=& - {\cal{G}}_{\theta}^{(5)}  
= -\frac{2}{9L^2}\left(\frac{L}{y_{I} + y_{0}}\right)^{\frac{8}{3}}, 
\eqn
where $ y_{1} = y_{c}> 0$ and $y_{2} = 0$. 
Inserting the above into Eqs.(\ref{4.14a}) and (\ref{4.14b}), and considering the
fact that $ H = 0$ we find that these two equations are satisfied for 
$\tau^{(I)}_{\mu\nu} = 0$, provided that the tension $\tau_{p}^{(I)}$
defined by Eq.(\ref{3.3b}) satisfies the relation, 
\bq
\lb{5.2}
\left(\tau_{(\phi,\psi_\pm)}^{(I)} + 2{\rho_\Lambda}^{(I)}\right)^2 = \frac{\rho_{\Lambda}^{(I)}}
      {9\pi G_4 L^2} \left(\frac{L}{y_I + y_0}\right)^{8/3}, 
\eq
where $\rho^{(I)}_{\Lambda}$ denotes the corresponding  energy density of the 
effective cosmological constant on the I-th brane, defined as
$\rho_{\Lambda}^{(I)} = \Lambda^{(I)}/(8\pi G)$.  
On the other hand, from Eqs.(\ref{4.14d}) and (\ref{4.14e}) we find that
\bqn
\lb{5.2ac}
\frac{\partial{V^{(I)}_{4}}}{\partial{\phi}} &=& 
\frac{c_1\epsilon_I}{\kappa_5^2(y_I + y_0)},\\
\lb{5.2ad}
\frac{\partial{V^{(I)}_{4}}}{\partial{\psi_{+}}} &=& 
\frac{c_2\epsilon_I}{\kappa_5^2(y_I + y_0)},\\ 
\frac{\partial{V^{(I)}_{4}}}{\partial{\psi_{-}}} &=& 
-\frac{\sqrt{3}c_2\epsilon_I}{\sqrt{2}\kappa_5^2(y_I + y_0)},
\eqn

To study the radion stability, it is found convenient to introduce the proper
distance $Y$, defined by 
\bq
\lb{5.2a}
Y = \frac{3L}{4} \left(\left(\frac{y_I + y_0}{L}\right)^{4/3} 
- \left(\frac{y_0}{L}\right)^{4/3}\right).
\eq
Then, in terms of $Y$, the static solution (\ref{5.1a}) can be written as
\bq
\lb{5.1aa}
ds^{2}_{5} = e^{-2A(Y)}\eta_{\mu\nu}dx^{\mu}dx^{\nu} - dY^{2},
\eq
with  
\bqn
\lb{5.1ab}
A(Y) &=& - \frac{1}{4}\ln\left(\frac{4\left(|Y| + Y_0\right)}{3L}\right),\nb\\
\phi(Y) &=& \frac{3}{4}c_1\ln\left(\frac{4\left(|Y| + Y_0\right)}{3L}\right) 
                + \phi_0,\nb\\
\psi_{+}(Y) &=&  \frac{3}{4} c_2\ln\left(\frac{4\left(|Y| + Y_0\right)}{3L}\right) 
              + \psi_{+}^0, \nb\\
\psi_{-}(Y) &=&  \sqrt{\frac{27}{32}}\; c_2\ln\left(\frac{4\left(|Y| + Y_0\right)}{3L}\right) 
+ \psi_{-}^0,
\eqn
where
\bq
\lb{5.1ac}
Y_{0} = \frac{3L}{4}\left(\frac{y_0}{L}\right)^{4/3}.
\eq 
Following \cite{GW99}, let us consider a massive scalar field $\Phi$ with the actions,
\bqn
\lb{5.3a}
S_{b} &=& \int{d^{4}x}\int_{0}^{Y_{c}}{dY \sqrt{-g_{5}} 
\left(\left(\nabla\Phi\right)^{2} - M^{2}\Phi^{2}\right)},\nb\\
S_{I} &=& - \alpha_{I} \int_{M^{(I)}_{4}}{d^{4}x \sqrt{-g_{4}^{(I)}} 
\left(\Phi^{2} - v^{2}_{I}\right)^{2}},
\eqn
where $\alpha_{I}$ and $v_{I}$ are real constants.  Then,  it can be shown 
that, in the background of Eq.(\ref{5.1aa}),  the massive scalar field $\Phi $ satisfies
the following Klein-Gordon equation 
\bq
\lb{5.5}
\Phi'' - 4A'\Phi' - M^{2}\Phi =
\sum_{I=1}^{2}{2  \alpha_{I} \Phi \left(\Phi^{2} - v^{2}_{I}\right)\delta(Y - Y_{I})}.
\eq
Integrating the above equation in the neighborhood of the I-th brane, we find that
\bq
\lb{5.5a}
\left. \frac{d\Phi(Y)}{dY}\right|_{Y_{I}-\epsilon}^{Y_{I}+\epsilon}
= 2\alpha_{I}  \Phi_{I} \left(\Phi^{2}_{I} - v^{2}_{I}\right),
\eq
where $\Phi_{I} \equiv \Phi(Y_{I})$.  Since
\bqn
\lb{5.5e}
\lim_{Y \rightarrow Y_{c}^{+}}{\frac{d\Phi(Y)}{dY}} &=& 
- \lim_{Y \rightarrow Y_{c}^{-}}{\frac{d\Phi(Y)}{dY}}   
\equiv - \Phi'\left(Y_{c}\right),\nb\\
\lim_{Y \rightarrow 0^{-}}{\frac{d\Phi(Y)}{dY}} &=& 
- \lim_{Y \rightarrow 0^{+}}{\frac{d\Phi(Y)}{dY}} \equiv - \Phi'(0),
\eqn
we find that the  conditions  (\ref{5.5a}) can be written in the forms,
\bqn
\lb{5.5fa}
\Phi'(Y_{c}) &=& - \alpha_{1}  \Phi_{1} \left(\Phi^{2}_{1} - v^{2}_{1}\right),\\
\lb{5.5fb}
\Phi'(0) &=& \alpha_{2}  \Phi_{2} \left(\Phi^{2}_{2} - v^{2}_{2}\right).
\eqn
Inserting the above solution back to the actions (\ref{5.3a}), and then integrating
them with respect to $Y$, we obtain the effective potential for the radion $Y_{c}$,
\bqn
\lb{5.5g}
 V_{\Phi}\left(Y_{c}\right) &\equiv&  - \int_{0+\epsilon}^{Y_{c}- \epsilon}{dY \sqrt{-g_{5}} 
\left(\left(\nabla\Phi\right)^{2} - M^{2}\Phi^{2}\right)}\nb\\
& & + \sum_{I=1}^{2}{ \alpha_{I}  \int_{Y_{I} -\epsilon}^{Y_{I} + \epsilon}
{dY \sqrt{-g_{4}^{(I)}} 
\left(\Phi^{2} - v^{2}_{I}\right)^{2}}}\nb\\
& & \;\;\;\;\;\;\; \;\;\;\;\;\;\; \times \delta\left(Y-Y_{I}\right) \nb\\
&=& \left. e^{-4A(Y)}\Phi(Y)\Phi'(Y)\right|^{Y_{c}}_{0}\nb\\
& &
+ \sum_{I=1}^{2}{\alpha_{I} \left(\Phi^{2}_{I} - v^{2}_{I}\right)^{2}e^{-4A(Y_{I})}}.
\eqn

For the background solution given by Eq.(\ref{5.1ab}), one find that in the region 
$ 0 < Y < Y_{c}$,  Eq.(\ref{5.5}) reads,
\bq
\lb{5.5h}
 \frac{d^{2}\Phi}{dz^{2}} + \frac{1}{z}\frac{d\Phi}{dz} - \Phi = 0,
\eq
where $z \equiv M\left(Y + Y_{0}\right)$. Eq.(\ref{5.5h}) has the general solution,
\bq
\lb{5.5i}
  \Phi = a I_{0}(z) + b K_{0}(z), 
\eq
where $I_{0}(z)$ and $K_{0}(z)$ denote the modified Bessel function of the first and
second kind, respectively \cite{AS72}. 	 
In the limit that $\alpha_{I}$'s are very large \cite{GW99}, Eqs.(\ref{5.5fa}) 
and (\ref{5.5fb}) show that there are solutions only when
$\Phi(0) \simeq v_{2}$ and $\Phi(Y_{c}) \simeq v_{1}$, that is, 
\bqn
\lb{5.5ha}
v_{1} &\simeq& a I_{0}^{c}  + b K_{0}^{c},\\
\lb{5.5hb}
v_{2} &\simeq&  a I_{0}^{0}  + b K_{0}^{0},
\eqn
where $z_{c} = M\left(Y_{c} + Y_{0}\right),\; z_{0} = M Y_{0},\; 
I^{i}_{0} \equiv I_{0}\left(z_{i}\right)$ and $K^{i}_{0} \equiv K_{0}\left(z_{i}\right)$. 
Eqs.(\ref{5.5ha}) and  (\ref{5.5hb}) have the solution,
\bqn
\lb{5.6a}
a &\simeq& \frac{1}{\Delta}\left(v_{1}K_{0}^{0}  - v_{2}K_{0}^{c}\right),\nb\\
b &\simeq&  \frac{1}{\Delta}\left(v_{2}I_{0}^{c}  - v_{1}I_{0}^{0}\right),
\eqn
where
\bq
\lb{5.7}
\Delta \equiv  I_{0}^{c}  K_{0}^{0}  - I_{0}^{0}  K_{0}^{c}.
\eq
Inserting Eqs.(\ref{5.5i}) and (\ref{5.6a}) into Eq.(\ref{5.5g}), we find that
\bqn
\lb{5.8}
V_{\Phi}\left(Y_{c}\right) &\simeq&  \frac{4}{3L\Delta}\left\{v_{1}z_{c}
        \left[v_{1}\left(I^{0}_{0}K^{c}_{1} + I^{c}_{1}K^{0}_{0}\right) \right.\right.\nb\\
	 & & \left.  - v_{2}\left(I^{c}_{0}K^{c}_{1} + I^{c}_{1}K^{c}_{0}\right)\right]\nb\\
 	& &   + v_{2}z_{0}
        \left[v_{2}\left(I^{c}_{0}K^{0}_{1} + I^{0}_{1}K^{c}_{0}\right) \right.\nb\\
	& & \left.\left.
	      - v_{1}\left(I^{0}_{0}K^{0}_{1} + I^{0}_{1}K^{0}_{0}\right)\right]\right\}.     
\eqn
To further study the potential, let us consider  two different limits, $z_{0} \gg 1$
and $z_{0} \ll 1$. With all these free parameters at hand, it is not difficult to see
that the mass of the radion should be also in the order of GeV, as we obtained
previously in both string \cite{WS08} and M theory \cite{WGW08}.

\subsection{$z_{0} \gg 1$}

\begin{figure}
\includegraphics[width=\columnwidth]{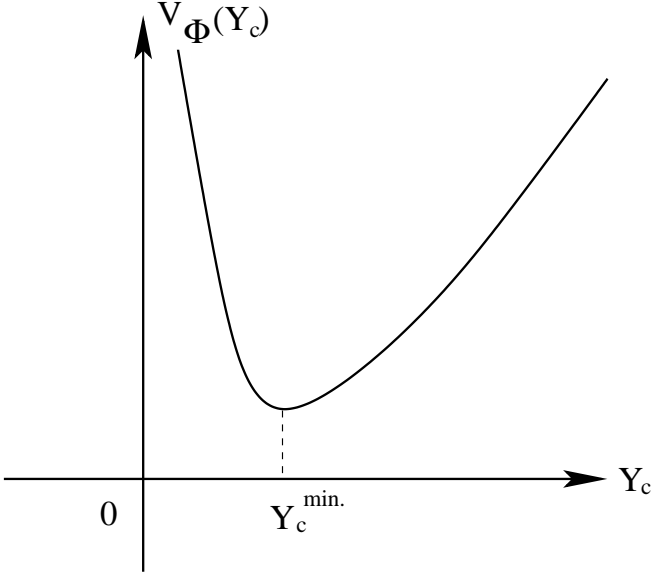}
\caption{The potential $V_{\Phi}\left(Y_{c}\right)$ given by Eq.(\ref{5.8a})
for $M \gg 1/Y_{0}$.   }
\label{fig3}
\end{figure}

When $z_{0} \gg 1$, we have $z_{c}  = z_{0} + MY_{c} \gg 1$. Then, we find
\bqn
\lb{5.9a}
I_0(z) &\simeq& I_1(z) \simeq \sqrt{\frac{1}{2\pi z}} e^z,\nb\\
K_0(z) &\simeq& K_1(z) \simeq \sqrt{\frac{\pi}{2 z}} e^{-z}. 
\eqn
Inserting the above expressions into Eq.(\ref{5.8}), we obtain
\bqn
\lb{5.8a}
V_{\Phi}\left(Y_{c}\right) &\simeq&  \frac{4z_{0}}{3L\sinh\left(MY_{c}\right)}
\left\{\left({v_1}^2  + {v_2}^2\right) \cosh\left(MY_{c}\right)  \right.\nb\\
& & \left.
  - 2  v_1 v_2  \right\},
\eqn
which has a minimum at
\bq
\lb{5.8b}
Y_{c}^{min.} = \frac{1}{M}\cosh^{-1}\left(\frac{{v_1}^2  + {v_2}^2}{2  v_1 v_2}\right),
\eq
where
\bqn
\lb{5.8c}
\left.\frac{\partial^{2}V_{\Phi}\left(Y_{c}\right)}{\partial{Y_{c}}^{2}} 
\right|_{Y_{c} = Y_{c}^{min.}}
&\simeq&  \left(\frac{16z_{0}M^{2}}{3L}\right)
\frac{\left(v_{1}v_{2}\right)^{2}}{\left|v^{2}_{1} - v^{2}_{2}\right|} > 0,\nb\\
V_{\Phi}\left(Y_{c}\right) &\simeq& \cases{\infty, & $Y_{c} = 0$,\cr
\infty, & $Y_{c} =  \infty$.\cr}
\eqn
Fig. \ref{fig3} shows the potential schematically, from which we can see that it
always has a minimum at a finite and non-zero value of $Y_{c}$. Therefore, in the 
present setup, the radion is stable in the limit $M \gg 1/Y_{0}$.

To calculate the corresponding 
radion mass, we need to know the precise relation between $Y_{c}$ and the radion
scalar $\varphi$. Following \cite{WGW08,GW99}, we find that
\bqn
\lb{5.8d}
\varphi &=& \left(\frac{12}{\kappa^{2}_{5}}\int_{0}^{Y_{c}}e^{-2A}dY\right)^{1/2}
= \sqrt{6 LM^{3}_{5}} \nb\\
&& \times \left\{\left(\frac{4\left(Y_{c} + Y_{0}\right)}{3L}\right)^{3/2} 
- \left(\frac{4Y_{0}}{3L}\right)^{3/2}\right\}^{1/2}.\;\;\;\;\;
\eqn
Then, we obtain that
\bqn
\lb{5.8e}
m^{2}_{\varphi} &\equiv& \left.  \frac{\partial^{2} V_{\Phi}\left(Y_{c}\right)}
{2\partial\varphi^{2}}\right|_{Y_{c} = Y^{min.}_{c}} = 
\frac{M^{2}}{M^{3}_{5}}\left(\frac{16Y_{0}}{27L}\right)^{1/2}
\nb\\
& & \times \frac{\left(v_{1}v_{2}\right)^{2}}{\left|v^{2}_{1} - v^{2}_{2}\right|}
 \cosh^{-1}\left(\frac{{v_1}^2  + {v_2}^2}{2  v_1 v_2}\right), 
\eqn
where $M^{3}_{5} = M^{8}_{10} V_{d_{+}}V_{d_{-}}$, as can be seen from Eqs.(\ref{2.2})
and (\ref{2.14a}).

\subsection{$z_{0} \ll 1$}

\begin{figure}
\includegraphics[width=\columnwidth]{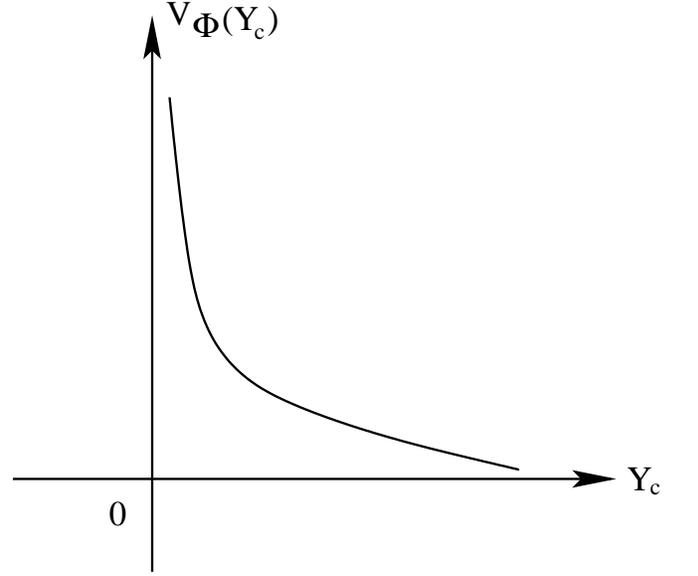}
\caption{The potential $V_{\Phi}\left(Y_{c}\right)$ given by Eq.(\ref{5.8f})
for $M \ll 1/Y_{0}$.   }
\label{fig4}
\end{figure}

When $z \ll 1$, we  find  
\bqn
\lb{5.9b}
I_0(z) &\simeq& 1,\;\;\;  
I_1(z) \simeq \frac{z}{4},\nb\\
K_0(z) &\simeq& - \ln(z),\;\;\;
K_1(z) \simeq \frac{1}{z}.
\eqn
Then, Eq.(\ref{5.8}) reduces to 
\bqn
\lb{5.8f}
V_{\Phi}\left(Y_{c}\right) &\simeq& \frac{v_{1}-v_{2}}{3LY_{c}}
\left\{\left(v_1 - v_2\right)\left(4 - z^{2}_{0} \ln\left(z_{0}\right)\right)
Y_{0}\right.\nb\\
& & \left.
+ z^{2}_{0}\left(v_{2} - 2v_{1}\ln\left(z_{0}\right)\right)Y_{c}\right\},
\eqn
for $Y_{c} \ll Y_{0}$. Fig. \ref{fig4}  shows the potential schematically, from which
 we can see that it has non-minimum. That is, the radion is not stable for $M \ll 1/Y_{0}$.
Combining it with last case, we find that there must exist a critical $M_{c}$, for which
the radion is stable when  $M > M_{c} > 0$, and not stable when $M < M_{c}$.

\section{ Localization of Gravity and 4D Effective Newtonian Potential}

\renewcommand{\theequation}{6.\arabic{equation}} \setcounter{equation}{0}

To study the localization of gravity and the four-dimensional effective
gravitational potential, in this section let us consider small fluctuations $%
h_{ab}$ of the 5-dimensional static metric with a 4-dimensional Poincar\'e
symmetry, given by Eqs.(\ref{5.1a}) in its conformally flat form.

\subsection{Tensor Perturbations and the KK Towers}

\begin{figure}[tbp]
\centering
\includegraphics[width=8cm]{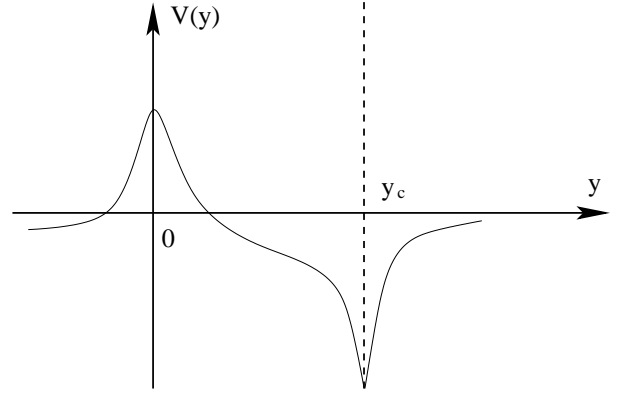}
\caption{The potential defined by Eq.(\ref{7.7}). }
\label{fig5}
\end{figure}

Since such tensor perturbations are not coupled with scalar ones \cite{GT00}%
, without loss of generality, we can set the perturbations of the scalar
fields to zero, i.e., $\delta \phi_{n} =0$. We
shall choose the gauge \cite{RS1,RS2}
\begin{equation}
h_{ay}=0,\;\;\;h_\lambda ^\lambda =0=\partial ^\lambda h_{\mu \lambda }.
\label{7.1}
\end{equation}
Then, it can be shown that \cite{Csaki00}
\begin{eqnarray}
\delta {G}_{ab}^{(5)} &=&-\frac 12\Box _5h_{ab}-\frac 32\left\{ \left(
\partial _c\sigma \right) \left( \partial ^ch_{ab}\right) \right.  \nonumber
\label{7.2} \\
&&\left. -2\left[\Box _5\sigma +\left( \partial _c\sigma \right) \left(
\partial ^c\sigma \right) \right] h_{ab}\right\} ,  \nonumber \\
\kappa _5^2\delta {T}_{ab}^{(5)} &=& -\frac{1}{4}h_{ab}
\left(\sum_{n=1}{\left(\nabla\phi_{n}\right)^{2}} - 2V_{5}\right),  \nonumber \\
\delta {T}_{\mu \nu }^{(4)} &=& \left(\tau_{p} + \lambda\right)h_{\mu\nu},  
\end{eqnarray}
where $\Box _5\equiv\eta ^{ab}\partial _a\partial _b$ and $\left( \partial
_c\sigma \right) \left( \partial ^ch_{ab}\right)\equiv \eta ^{cd}\left(
\partial _c\sigma \right) \left( \partial _dh_{ab}\right) $, with $\eta
^{ab} $ being the five-dimensional Minkowski metric. Substituting the above
expressions into the gravitational field equations (\ref{3.3}) with $D=5$, 
we find that in the present case there is only one independent equation,
given by 
\begin{equation}
\Box _5\tilde h_{\mu \nu }+\frac 32\left(\sigma ^{\prime \prime }+\frac
32{\sigma ^{\prime }}^{2}\right) \tilde h_{\mu \nu }=0,  \label{7.4}
\end{equation}
where $h_{\mu \nu }\equiv e^{-3\sigma /2}\tilde h_{\mu \nu }$. Setting
\begin{eqnarray}
&&\tilde h_{\mu \nu }(x,y)=\hat h_{\mu \nu }(x)\psi (y),  \nonumber
\label{7.5} \\
&&\Box _5=\Box _4 - \nabla _y^2 =\eta ^{\mu \nu
}\partial _\mu \partial _\nu - \partial _y^2,  \nonumber \\
&&\Box _4\hat h_{\mu \nu }(x)= - m^2\hat h_{\mu \nu }(x),
\end{eqnarray}
we find that Eq.(\ref{7.4}) takes the form of the schr\"odinger equation,
\begin{equation}
\left( -\nabla _y^2+V\right) \psi =m^2\psi ,  \label{7.6}
\end{equation}
where 
\begin{eqnarray}
V &\equiv &\frac 32\left( \sigma ^{\prime \prime }+\frac 32{\sigma ^{\prime }%
}^2\right)  \nonumber  \label{7.7} \\
&=& - \frac{1}{4\left(\left|y\right| + y_{0}\right)^{2}} 
+ \frac{\delta(y)}{y_{0}} - \frac{\delta\left(y - y_{c}\right)}{y_{c} + y_{0}}.
\end{eqnarray}
From the above expression we can see clearly that the potential has a
delta-function well at $y=y_c$, which is responsible for the localization of
the graviton on this brane. In contrast, the potential has a delta-function
barrier at $y=0$, which makes the gravity delocalized on the $y=0$ brane.
Fig. \ref{fig5} shows the potential schematically.

Integration of Eq.(\ref{7.6}) in the neighbourhood of 
$y=0$ and $y=y_c$ yields, respectively, the boundary conditions,
\begin{eqnarray}
 \label{7.12a}
\lim_{y\rightarrow y_c^{-}}{\psi ^{\prime }(y)} &=&\frac{1}
{2\left(y_c+y_0\right)}\lim_{y\rightarrow y_c^{-}}{\psi (y)},  \\
 \label{7.12b}
\lim_{y\rightarrow 0^{+}}{\psi ^{\prime }(y)} &=&\frac{1}{2y_0}
\lim_{y\rightarrow 0^{+}}{\psi (y)}. 
\end{eqnarray}
Note that in writing the above equations we had used the $Z_2$ symmetry of
the wave function $\psi $. 

Introducing the operators,
\begin{equation}
Q\equiv \nabla _y-\frac 32\sigma ^{\prime },\;\;\;Q^{\dagger }\equiv -\nabla
_y-\frac 32\sigma ^{\prime },  \label{7.8}
\end{equation}
Eq.(\ref{7.6}) can be written in the form of a supersymmetric quantum
mechanics problem, 
\begin{equation}
 \label{7.9}
Q^{\dagger }\cdot Q\psi =m^2\psi,
\end{equation}
which, together with the boundary conditions (\ref{7.12a}) and (\ref{7.12b}),  
guarantees that the
operator $Q^{\dagger }\cdot Q$ is Hermitian \cite{Csaki01, WGW08}.  Then, by 
the usual theorems from Quantum Mechanics \cite{QM}, we can see
that all eigenvalues $m^2$ are non-negative, and their corresponding wave
functions $\psi _n(y)$ are orthogonal to each other and form a complete
basis. Therefore, {the background in the current setup is gravitationally stable}.

\subsubsection{Zero Mode}

\begin{figure}[tbp]
\centering
\includegraphics[width=8cm]{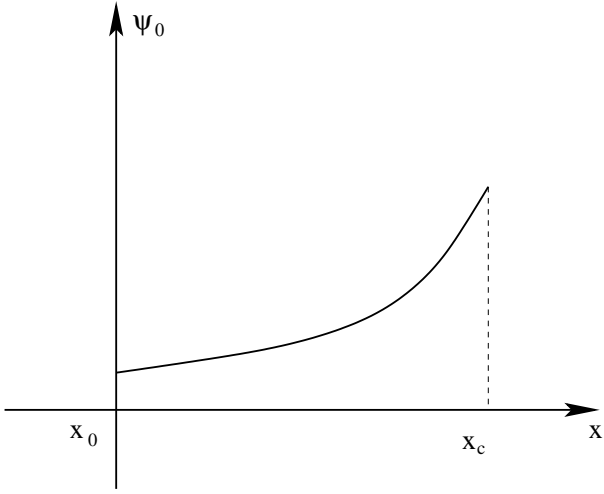}
\caption{The zero-mode wavefunction given by Eq.(\ref{7.10}), from which we can see
that the gravity is localized on the visible brane at $x = x_{c}$. }
\label{psi0}
\end{figure}

The four-dimensional gravity is given by the existence of the normalizable
zero mode, for which the corresponding wavefunction is given by
\begin{equation}
\psi _0(y)= N_{0}\left(\left|y\right| + y_{0}\right)^{1/2},  \label{7.10}
\end{equation}
where $N_0$ is the normalization factor, defined as
\begin{equation}
N_0 = \sqrt{\frac{2}{y_{c}\left(y_{c} + 2y_{0}\right)}}.
\label{7.11}
\end{equation}
Eq.(\ref{7.10}) shows clearly that the wavefunction is increasing as $y$
increases from $0$ to $y_c$ [cf. Fig. \ref{psi0}]. Therefore, the gravity is 
indeed localized near the $y=y_c$ brane.

\subsubsection{Non-Zero Modes}

\begin{figure}[tbp]
\centering
\includegraphics[width=8cm]{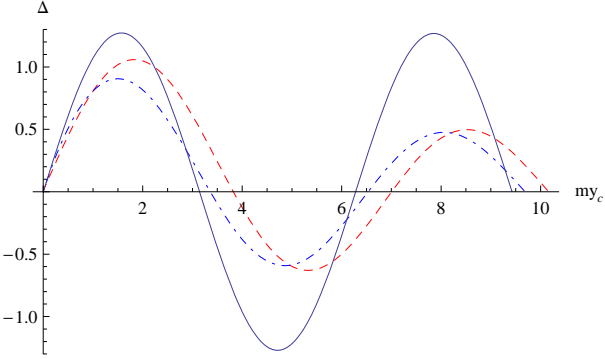}
\caption{The re-scaled function of $\Delta$ defined by Eq.(\ref{7.18}), where 
the dashed, dot-dashed and solid lines are,  respectively,  for 
$\Delta\left(x_{0} = 0.01\right)/35; \Delta\left(x_{0} = 1.0\right)/0.5$
and $\Delta\left(x_{0} =1000\right)/0.005$.  }
\label{Delta}
\end{figure}

In order to have localized four-dimensional gravity, we require that the
corrections to the Newtonian law from the non-zero modes, the KK modes, of
Eq.(\ref{7.6}), be very small, so that they will not lead to contradiction
with observations. When $m \not= 0$, it can be shown that Eq.(\ref{7.6})
has the general solution,
\begin{equation}
\label{7.15}
\psi =x^{1/2}\left(cJ_{0}(x)+ dY_{0}(x)\right), 
\end{equation}
where $x \equiv m\left(y + y_{0}\right)$, and $J_{0}(x)$ and $Y_{0}(x)$
are the Bessel functions of the first and second kind, respectively \cite{AS72}.
The integration constants $c$ and $d$ are   determined from the boundary conditions,
 Eqs.(\ref{7.12a}) and (\ref{7.12b}), which  can now be cast in the form,
\begin{equation}
\left(\matrix{J_{1}\left(x_{c}\right) & Y_{1}\left(x_{c}\right) \cr
J_{1}\left(x_{0}\right) & Y_{1}\left(x_{0}\right) \cr}
\right)\left( \matrix{c \cr d\cr}\right) =0,  \label{7.17}
\end{equation}
where $x_{0} \equiv my_{0}$ and $x_{c} \equiv x_{0} + my_{c}$.
Clearly, it has no trivial solutions only when
\bqn
\label{7.18}
\Delta\left(x_{0}, x_{c}\right) &\equiv&
J_{1}\left(x_{c}\right)Y_{1}\left(x_{0}\right)
- J_{1}\left(x_{0}\right)Y_{1}\left(x_{c}\right)\nb\\
&=& 0.  
\eqn
Fig. \ref{Delta} shows the function $\Delta\left(x_{0}, my_{c}\right)$ 
for $x_0=0.01,\;1.0,\;1000$, respectively. Note that in plotting these
lines, properly rescaling toke place. From this figure, we find that 
the spectrum of the gravitational KK towers is discrete, and 
weakly depends on the specific values of $x_0$.

\begin{table}[tbp]
\begin{tabular}{|c|c|c|c|}
\hline
\label{table1} $x_{0}$ & $m_{1}y_{c}$ & $m_{2}y_{c}$ & $m_{3}y_{c}$ \\ \hline
0.01 & 3.82 & 7.01 & 10.16 \\ \hline
1.0 & 3.36 & 6.53 & 9.69 \\ \hline
1000 & 3.14 & 6.28 & 9.42 \\ \hline
\end{tabular}
\caption{The first three modes $m_{n}\; (n = 1, 2, 3)$ for $x_{0} = 0.01, \;
1.0, \; 1000$, respectively.}
\end{table}

Table I shows the first three modes $m_{n}\; (n = 1, 2, 3)$ for $x_{0} =
0.01, \; 1.0, \; 1000$, from which we can see that to find $m_{n}$ it is
sufficient to consider only the case where $x_{0} \gg 1$.
When $x_0\gg 1$ we find that $x_c = x_0+my_c\gg 1$ and \cite{AS72}
\begin{eqnarray}
\label{7.19}
J_{1}(x) &\simeq & \sqrt{\frac{2}{\pi x}}\cos\left(x - \frac{3}{4}\pi\right),
   \nonumber   \\
Y_{1}(x) &\simeq & \sqrt{\frac{2}{\pi x}}\sin\left(x - \frac{3}{4}\pi\right).
\end{eqnarray}
Inserting the above expressions into Eq.(\ref{7.18}), we
obtain
\begin{eqnarray}
\Delta  &=& \frac{2}{\pi\sqrt{x_{0}x_{c}}}\sin\left(m y_{c}\right),
\end{eqnarray}
whose roots are given by
\begin{equation}
m_{n} = \frac{n\pi}{y_{c}}, \;\;\; (n = 1, 2, ...).
\label{7.21}
\end{equation}
In particular, we have
\begin{eqnarray}
m_1 &\simeq &3.14\times \left( \frac{10^{-19}\;{\mbox{m}}}{y_c}\right) \;{%
\mbox{TeV}}  \nonumber  \label{7.24} \\
&\simeq &\cases{1\; {\mbox{TeV}}, & $y_{c} \simeq 10^{-19} \;
{\mbox{m}}$,\cr 10^{-2} \; {\mbox{eV}}, & $y_{c} \simeq 10^{-5} \;
{\mbox{m}}$,\cr 10^{-4} \; {\mbox{eV}}, & $y_{c} \simeq 10^{-3} \;
{\mbox{m}}$.\cr}
\end{eqnarray}

It should be noted that the   mass $m_{n}$ calculated above is measured
by the observer with the metric $\eta_{\mu\nu}$. However, since the warped factor
$e^{\sigma(y)}$ is not one at $y = y_{c}$, the physical mass on the visible brane 
should be given by \cite{RS1}
\bq
\lb{phycialmass}
m^{obs}_{n} = e^{-\sigma\left(y_{c}\right)}m_{n}
= \left(\frac{y_{c} + y_{0}}{L}\right)^{1/3} m_{n}.
\eq
Without introducing any new hierarchy, we expect that 
$\left[({y_{c} + y_{0})}/{L}\right]^{1/3} \simeq {\cal{O}}(1)$. As a result, 
we have 
\bq
\lb{phycialmassb}
m^{obs}_{n}  
= \left(\frac{y_{c} + y_{0}}{L}\right)^{1/5} m_{n} \simeq m_{n}.
\eq

For each $m_{n}$ that satisfies Eq.(\ref{7.18}), the wavefunction $%
\psi_{n}(y)$ is given by
\begin{eqnarray}  \label{7.25}
\psi_{n}(y) &=& N_{n}x_{n}^{1/2}\left(\frac{J_{0}\left(x_{n}\right)}
 {J_{1}\left(x_{0, n}\right)} 
- \frac{Y_{0}\left(x_{n}\right)}{Y_{1}\left(x_{0, n}\right)}\right),    
\end{eqnarray}
where 
\bqn
\lb{7.25a}
x_{0,n} &\equiv& m_{n} y_{0} \simeq n\pi \left(\frac{y_{0}}{y_{c}}\right),\nb\\
x_{n} &\equiv& m_{n} \left(y_{0} + y\right) \simeq n\pi \left(\frac{y_{0} + y}{y_{c}}\right).
\eqn
The normalization factor $N_{n} [\equiv N_{n}\left(m_{n}, y_{c}\right)]$ is  
determined by the condition,
\begin{equation}  
\label{7.26}
\int^{y_{c}}_{0}{\left|\psi_{n}(y)\right|^{2} dy} = 1.
\end{equation}

\begin{figure}[tbp]
\centering
\includegraphics[width=8cm]{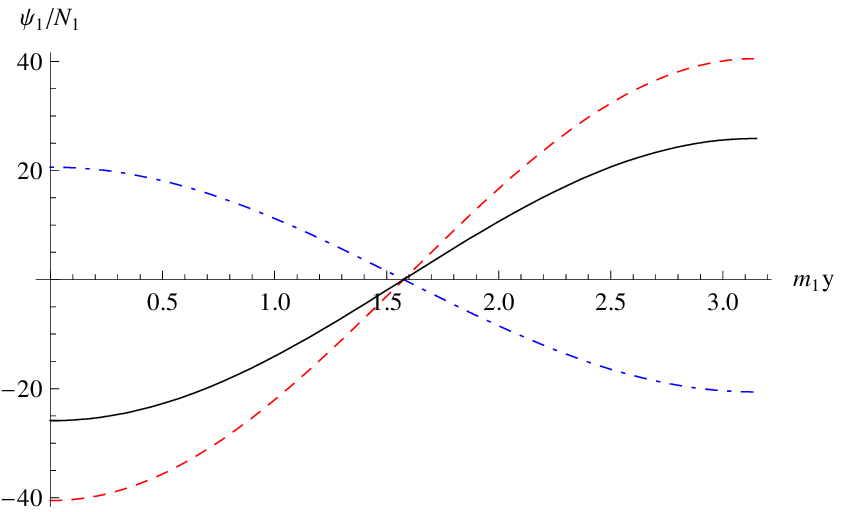}
\caption{The  wavefunction, $\psi_{1}\left(y\right)$,
defined by Eq.(\ref{7.25}) vs $m_{1}y$ where $ y \in \left[0, y_{c}\right]$.
The dashed, dot-dashed and solid lines are,  respectively,  for 
$x_{0, 1} = 100, \; 102,\; 104$.  }
\label{psi1}
\end{figure}

\begin{figure}[tbp]
\centering
\includegraphics[width=8cm]{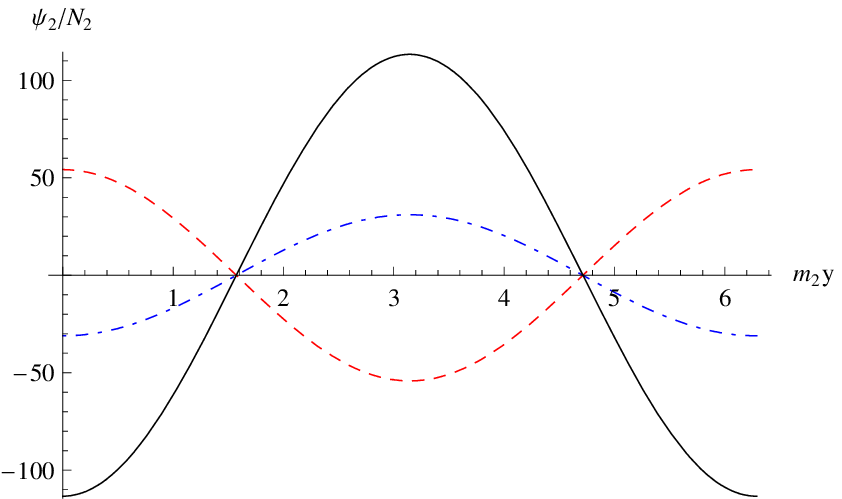}
\caption{The  wavefunction, $\psi_{2}\left(y\right)$,
defined by Eq.(\ref{7.25}), vs $m_{2}y$ where $ y \in \left[0, y_{c}\right]$.
The  dashed, dot-dashed and solid lines are,  respectively,  for 
$x_{0, 1} = 100, \; 102,\; 104$.  }
\label{psi2}
\end{figure}

\begin{figure}[tbp]
\centering
\includegraphics[width=8cm]{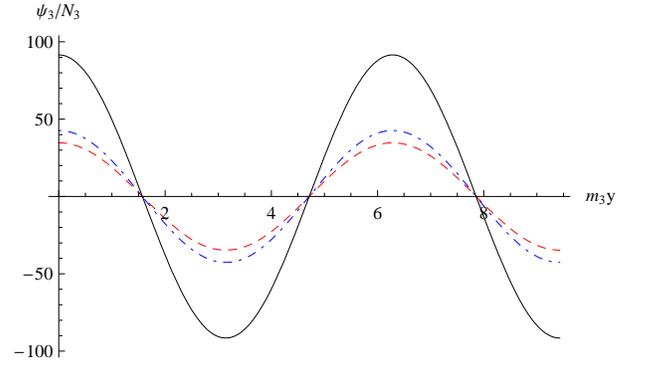}
\caption{The  wavefunction, $\psi_{3}\left(y\right)$,
defined by Eq.(\ref{7.25}), vs $m_{3}y$ where $ y \in \left[0, y_{c}\right]$.
The   dashed, dot-dashed and solid lines are,  respectively,  for 
$x_{0, 1} = 100, \; 102,\; 104$.  }
\label{psi3}
\end{figure}

Figs. \ref{psi1}, \ref{psi2} and \ref{psi3} show $\psi_{1}\left(y\right),\; 
\psi_{2}\left(y\right)$ and $\psi_{3}\left(y\right) $ for 
$x_{0, 1} = 100, \; 102, \; 104$, respectively.

\subsection{4D Newtonian Potential and Yukawa Corrections}

To calculate the four-dimensional effective Newtonian potential and its
corrections, let us consider two point-like sources of masses $M_1$ and $M_2$, 
located on the brane at $y=y_c$. Then, the discrete eigenfunction $\psi
_n(z)$ of mass $m_n$ has an Yukawa correction to the four-dimensional
gravitational potential between the two particles \cite{BS99,Csaki00},
\begin{equation}
U(r)=G_4\frac{M_1M_2}r+\frac{M_1M_2}{M_5^3r}\sum_{n=1}^\infty {%
e^{-m_nr}\left|\psi _n\left(y_{c}\right)\right| ^2},  \label{7.27}
\end{equation}
where $\psi_n\left(y_{c}\right)$ is given by Eq.(\ref{7.25}), with   
\bq
\lb{7.28}
x_{c,n} \equiv    m_{n}\left(y_{c} + y_{0}\right) 
\simeq \frac{n\pi y_{0}}{y_{c}} + n\pi.
\eq
When $x_{0,1} = m_{1}y_{0} \gg 1$, we find that 
\bqn
\lb{7.29}
N_{n} &\simeq& \frac{\cos\left(2m_{n}y_{0}\right)}{\sqrt{2n\pi y_{0}}},\nb\\
\psi_{n}\left(y_{c}\right) &\simeq& (-1)^{n+1}\sqrt{\frac{2}{y_{c}}}.
\eqn
Then, we obtain,
\bq
\lb{7.30}
\left|\psi _n\left(y_{c}\right)\right|^2 \simeq 2M_{pl}
\left(\frac{l_{pl}}{y_{c}}\right).
\eq
Clearly, by properly choosing $y_{c}$, the corrections of the 4-dimensional
Newtonian potential due to the high order gravitational KK modes are negligible.

\section{Conclusions} 

In this paper,  we have systematically studied the possibility of implementing 
the RS1 scenario \cite{RS1} into  type II string theory on an $S^{1}/Z_{2}$ orbifold. 
In particular, in Sec. II, starting with the Neveu-Schwarz/Neveu-Schwarz (NS/NS) 
sector, we have first compactified  the $\left(D+d_{+} + d_{-}\right)$-dimensional 
spacetime on two manifolds $M_{d_{+}}$ and $M_{d_{-}}$, where the topologies of 
$M_{d_{+}}$ and $M_{d_{-}}$ are unspecified.  As shown explicitly there, this 
particularly allows the dilaton and modulus fields to have non-zero potentials 
(masses), which is in contrast to the  toroidal compactification considered 
previously \cite{WS07,WSVW08,WS08,TW09,LWC00,BW06}.  After reducing the action 
to an effective $D$-dimensional one, which is given by Eq.(\ref{2.13}) in the 
Einstein frame,  we further compactify one of the $(D-1)$ spatial dimensions 
on an $S^{1}/Z_{2}$ orbifold, by adding the brane actions (\ref{3.1}). This
completes the whole setup of the model to be studied in this paper.  Lifting
it to the original spacetime, the two orbifold branes become 
($D+d_{+} + d_{-}- 1$)-dimensional.

In Sec.III, we have explicitly derived the corresponding gravitational and matter 
field equations both in the bulk and on the branes, by using the Gauss-Codacci
and Lanczos equations.  In Sec. IV such developed formulas have been applied to 
cosmology by setting $D = 5 = d_{+} + d_{-}$. In particular, the generalized 
Friedmann equations  on the branes are given explicitly by Eqs.(\ref{4.14a})
and (\ref{4.14b}).

In Sec. V, in order to study the radion stability and radion mass, we have first
derived the general static solutions with a 4-dimensional Poincar\'e symmetry.
Then, using the Goldberger-Wise mechanism, we have studied the radion stability
and shown explicitly that it is indeed stable in our current setup. The
corresponding radion mass is given by Eq.(\ref{5.8e}), from which we can see that
the observational constraint $m_{\varphi} > 10^{-3} \; eV$ can be easily satisfied
by properly choosing the free parameters presented in the model. 

In Sec. VI, we have studied the tensor perturbations, and shown explicitly that
the background solution is gravitational stable, and the gravity is localized 
on the visible brane, as one can be seen clearly from Fig. \ref{psi0}. Due to the
particular boundary conditions, the spectrum of the gravitational KK towers is
discrete, and the corresponding masses can be well approximated by Eq.(\ref{7.21}),
as one can see from Fig. \ref{Delta} and Table I. The mass gap $\Delta m \equiv m_{1}$
between the ground state and the first excited state can be in the order of $TeV$,
while the high order Yukawa corrections to the 4-dimensional Newtonian potential, 
due to the high order KK modes, is exponentially suppressed, and can be 
negligible.   

The above results strongly support our earlier conclusions obtained in the studies
of orbifold branes in both the HW heterotic M theory \cite{GWW07,WGW08} and 
string theory \cite{WS07,WSVW08,WS08}. In particular, in all these models the radion
is stable, and the gravity is localized on the visible (TeV) branes, in contrast to the
RS1 model \cite{RS1}, where the gravity is localized on the invisible brane. 
Our models are much more complicated than the RS1 model and involve several free
parameters. By properly choosing them,  the theory should be consistent with  observational 
constraints,  a subject that is under our
current investigations. It would be also extremely interesting to find specific
models in the current setup to explain the late cosmic acceleration of the universe
\cite{DEs}.

\section*{Acknowledgments}

One of the authors (AW) would like to 
 thank  K. Koyama,  R. Maartens, A. Papazoglou, Y.-S. Song and D. Wands for valuable discussions. 
 He also would also like to express his gratitude to the Institute 
 of Cosmology and Gravitation (ICG) for hospitality. 
This work was partially supported  by NSFC under grant No. 10703005 and No. 10775119  
 (AW $\&$ QW).

\end{document}